\documentclass[useAMS,usenatbib]{mn2e}

\usepackage{graphicx}
\usepackage{bm}
\usepackage{lipsum}
\usepackage{color}
\newcommand\fo{f_{\rm 1O}}
\newcommand\fmod{f_{\rm m}}

\title[Blazhko Effect in RRc stars]{Blazhko Effect in the first overtone RR Lyrae stars of the OGLE Galactic bulge collection}
\author[Netzel, et al.]
{H. Netzel$^{1}$\thanks{E-mail: henia@netzel.pl},
R. Smolec$^{1}$,
I. Soszy\'nski$^{2}$,
A. Udalski$^{2}$\\
$^{1}$Nicolaus Copernicus Astronomical Centre, Polish Academy of Sciences, Bartycka 18, 00-716 Warszawa, Poland\\
$^{2}$Warsaw University Observatory, Al. Ujazdowskie 4, 00-478 Warszawa, Poland\\
}

\begin{document}

\date{Accepted . Received ; in original form }

\pagerange{\pageref{firstpage}--\pageref{lastpage}} \pubyear{2017}

\maketitle

\label{firstpage}

\begin{abstract}
  We present the analysis of the Blazhko effect -- quasi-periodic modulation of pulsation amplitude and/or phase -- in the Galactic bulge first overtone RR~Lyrae stars (RRc). We used the data gathered during the fourth phase of the Optical Gravitational Lensing Experiment (OGLE). Out of $10\,826$ analyzed RRc stars, Blazhko effect was detected in  607 stars which constitute 5.6 percent of the sample. It is the largest and most homogeneous sample of modulated RRc stars analyzed so far. Modulation periods cover a wide range, from slightly above 2\,d to nearly 3000\,d. Multiperiodic modulation was detected in 47 stars. The appearance of modulation in the frequency domain was studied in detail. Modulation manifests either as close doublets or as equidistant triplets and multiplets centered on radial mode frequency and its harmonics. In a significant fraction (29 percent) of stars, we have detected the modulation frequency itself, which corresponds to the modulation of the mean stellar brightness. Our search for period doubling effect, that was discovered recently in modulated fundamental mode RR~Lyrae stars, and triggered development of new model behind the Blazhko modulation, yielded negative result. In 104 stars we detected additional signals that could correspond to both radial and non-radial modes. Statistical properties of modulated stars were analyzed in detail and confronted with properties of non-modulated stars and of modulated fundamental mode RR~Lyrae stars. Our analysis provides constraints for the models to explain the Blazhko phenomenon, which still remains a puzzle more than hundred years after its discovery.
\end{abstract}


\section{Introduction}\label{sec.intro}
RR~Lyrae stars are classically pulsating stars. They pulsate mostly in radial fundamental mode (RRab stars), radial first overtone (RRc stars) or in both modes simultaneously (RRd stars). We also know RR~Lyrae stars pulsating in radial fundamental mode and radial second overtone, but these stars are very rare \citep[e.g.][]{benko1,benko2}. Pulsations of RR~Lyrae stars have large amplitudes (up to $1$\,mag in the $I$ band) and periods from around $0.2$\,d to $1$\,d.

RR~Lyrae stars are important in many aspects of astrophysics. They serve as excellent distance indicators and tracers of old stellar populations. They are used in studies of kinematic and physical features of Galaxy and other stellar formations \citep[e.g.][]{csbook}. Despite their importance and relative simplicity of their pulsation, there are still problems concerning them which are not fully understood. The most stubborn problem is the Blazhko phenomenon. Blazhko effect is quasi-periodic modulation of amplitude and/or phase of a subgroup of RR~Lyrae stars. It was discovered in 1907 when Sergei Nicolaevich Blazhko observed quasi-periodic changes in moments of maxima in the light curve of RRab star RW~Dra \citep{blazhko}. Later, in 1916, Harlow Shapley \citep{shapley} detected changes of shape of light curve in RR~Lyr, the prototype of the class. 

Our knowledge about the Blazhko effect significantly increased in recent years thanks to both ground and space observations. Large-scale surveys, like Optical Gravitational Lensing Experiment \citep[OGLE,][]{ogle}, or the MAssive Compact Halo Objects \citep[MACHO, e.g.][]{alcockrrc}, have contributed to the discovery of thousands of RR~Lyrae stars in the Magellanic Clouds and in the Galactic bulge. Long time base of observations allowed to discover and to study the Blazhko effect in a significant fraction of these stars \citep[e.g.][]{alcockrrc,alcockrrab,nagykovacs,mizerski,rrabbl}. All studies show that the incidence rate of the Blazhko phenomenon is higher for RRab stars than for RRc and can exceed 50\,per cent \citep[e.g.][]{jurcsik2,benko2,ngc}. In RRc stars the reported incidence rates rarely exceed 10\,per cent -- see Sect.~\ref{ssec:incrate} for a more detailed discussion. Blazhko effect was also detected in a significant fraction of anomalous RRd stars \citep{soszrrd}.

Several breakthrough results emerged  when precise and nearly continuous space photometry gathered by {\it CoRoT} and {\it Kepler} became available. The most important discovery was detection of the period doubling effect in several Blazhko RRab stars, most notably in the eponym of the class, RR~Lyr \citep{rrlyrpd,szabo_doubling,benko1,szabo_doubling2}. Period doubling manifests as alternating deep and shallow brightness maxima during some phases of the Blazhko cycle. So far the effect was detected only in the modulated RRab stars. Space photometry allowed detailed studies of the light curve changes during the Blazhko cycle \citep[e.g.][]{kk_rrlyr,guggenberger2012}. Stability of the modulation cycles could be studies as well, clearly showing that the Blazhko effect might be strongly irregular phenomenon \citep[e.g.][]{benko2,guggenberger2011}. Interesting results concern the appearance of low-amplitude additional modes in modulated stars \citep{benkoszabo}.

Despite the fact that a significant fraction of RR~Lyrae stars seems to be modulated, there is no satisfactory model which explains all observed features of the Blazhko effect -- see \cite{kovacs_aip,kovacs_cokon} for recent reviews. Most theoretical work focused on RRab stars. The recent work was triggered by the discovery of the period doubling effect in the modulated RRab stars. \cite{92resonance} connected the two effects, Blazhko modulation and period doubling, and proposed that they are caused by the same mechanism: resonant interaction between the fundamental mode and the radial ninth overtone. While this model needs a verification through hydrodynamic modeling, its application to Blazhko RRc stars is not straightforward. First, the period doubling effect was not detected in Blazhko RRc stars so far; second, the suitable half-integer resonance that could work in first overtone pulsators was not proposed.

Admittedly, most of the studies concerning the Blazhko effect is focused on RRab stars. No modulated RRc star was observed by {\it CoRoT} or by {\it Kepler} in its original Cygnus field. Most of the ground-based studies were also focused on RRab pulsators. In this paper we focus on Blazhko RRc stars and aim to provide comprehensive study of the effect based on a large and homogeneous sample of stars. The fourth phase of OGLE project regularly monitors the Galactic bulge from 2010 \citep{o4}. Its excellent photometry, long time-base of data, and large number of objects, make it perfect for study of the Blazhko effect.

\section{Data and analysis}\label{sec.analysis}
In the OGLE collection of variable stars there are over $38\,000$ RR~Lyrae stars in the Galactic bulge \citep{o4bulge}. Among them, $10\,826$ are of RRc type\footnote{While the present analysis was finished, the OGLE collection of variable stars was updated, \citep{sosz_morerrl}, and currently counts 11415 RRc stars}. We analyzed $I$-band photometry of these stars to find the modulation of pulsation amplitude and phase, the Blazhko effect.

In the frequency spectrum, Blazhko effect manifests as equally spaced multiplets centered on main pulsation frequency, $\fo$, and its harmonics, $k\fo$. The separation between multiplet components corresponds to the modulation frequency, $\fmod$. Amplitude of the modulation sidepeaks depends on the modulation properties, e.g. phase relation between amplitude and phase modulation \citep[see, e.g.][]{bsp11}. Strongly asymmetric multiplets are possible. In the ground-based photometry, due to relatively large noise levels, we can detect mostly triplets, i.e. peaks located at $k\fo\pm\fmod$, or dublets. In the latter case, the modulation sidepeaks are located only on one side of $k\fo$.

Multiplets, including incomplete ones, and triplets, are commonly interpreted as due to modulation. In the following, stars with such structures in the frequency spectrum will be denoted as BLm. Dublets are also considered to be signature of modulation in RR~Lyrae stars. We note however, that modulation is not the only possible explanation of the observed signals. Frequency spectrum in which additional sidepeaks appear only on one side of $k\fo$, e.g. at $k\fo+\fmod$, can be interpreted as arising due to excitation of additional mode and its linear frequency combinations. This problem will be discussed in Sec.~\ref{sec.discussion} in more detail. In the following stars in which we detect the doublets will be denoted as BL1.

In some cases, we can detect a signal at the modulation frequency, $\fmod$, which arises due to modulation of the mean brightness.

We used standard consecutive prewhitening method in order to search for these characteristic features in the power spectra of RRc stars. Large number of stars motivated the development of automatic procedure dedicated to analyze RRc stars and to select candidates for further, more detailed analysis.

The first step of the automatic procedure is to prewhiten spectra with a frequency of the first overtone, $\fo$, and its harmonics, $k\fo$. To this aim, the Fourier series of the form:
$$ m(t)=A_0+\sum A_k \sin(2\pi k\fo t+\phi_k) $$
is fitted using non-linear least-square method and subtracted from the data. Amplitudes, phases and frequencies are adjusted during the fitting procedure. Possible trend present in the residual data is removed using a third order polynomial.

After first prewhitening and trend removal the automatic procedure checks the power spectrum for features characteristic for the Blazhko effect, i.e. multiplets around $\fo$ and its harmonics. The inspected range is $(\fo\pm 0.5)$\,d$^{-1}$, excluding the very close vicinity of $\fo$, as described in the next paragraph. The highest frequency in this range is automatically fitted to the data if signal to noise ratio (S/N) for the corresponding peak exceeds 4. Signal to noise ratio is determined from the power spectrum, where the noise is estimated as a mean amplitude in the 0-10 c/d range. The frequency separation between the first overtone frequency and the highest signal from its vicinity (presumed modulation frequency) is denoted by $\fmod$. The algorithm checks whether there are other structures in the spectrum characterized by the same separation. First it checks whether the sidepeak on the other side of $\fo$ is present. Then it checks higher-order harmonics for signals with the same separation, $\fmod$, and it checks low frequency range for the signal at modulation frequency, $\fmod$. Only signals with S/N above 4 are fitted to the data. In addition, after fitting a new frequency, the procedure checks whether it fulfills the criterion $A/\sigma \ge 4$, where $A$ is the amplitude of the fitted signal and $\sigma$ is its error. Only frequencies fulfilling these requirements are included in the final fit in the form $k\fo\pm\fmod$. Last step of the procedure was data clipping with $4\sigma$ criterion.

In the procedure outlined above, a close vicinity of the first overtone frequency is not inspected for the presence of additional signals. It is given by $\fo\pm 2/T$, in which $T$ is data length, and is related to the resolution of the Fourier transform. Formal resolution of the Fourier transform is $1/T$, but in this analysis we adopted a more conservative criterion and considered two frequencies to be resolved only if the separation between them is larger than $2/T$. As a result, to claim the modulation, we require that at least two modulation cycles are covered. Another consequence is an upper limit on the modulation period that can be detected in the data, that depends on their length, and may differ from star to star.

All OGLE RRc stars from the Galactic bulge were analyzed by this automatic procedure. This analysis resulted in 3066 Blazhko candidates (28\,per cent). Star was considered to be a Blazhko candidate when in its power spectrum at least two sidepeaks were detected at $k\fo$ with the same separation.

The automatic procedure has to deal with additional signals present in the power spectra besides $\fo$, its harmonics, and potential Blazhko structures. In power spectra of some stars there are instrumental signals (most often at $\approx\!1\,{\rm d}^{-1}$ or $\approx\!2\,{\rm d}^{-1}$), manifestations of additional modes,  of period changes and of remnant trends. All these signals are accompanied by their daily and annual aliases caused by regular gaps in the data.

\begin{figure}
\centering
\resizebox{\hsize}{!}{\includegraphics[bb= 0 0 570 473]{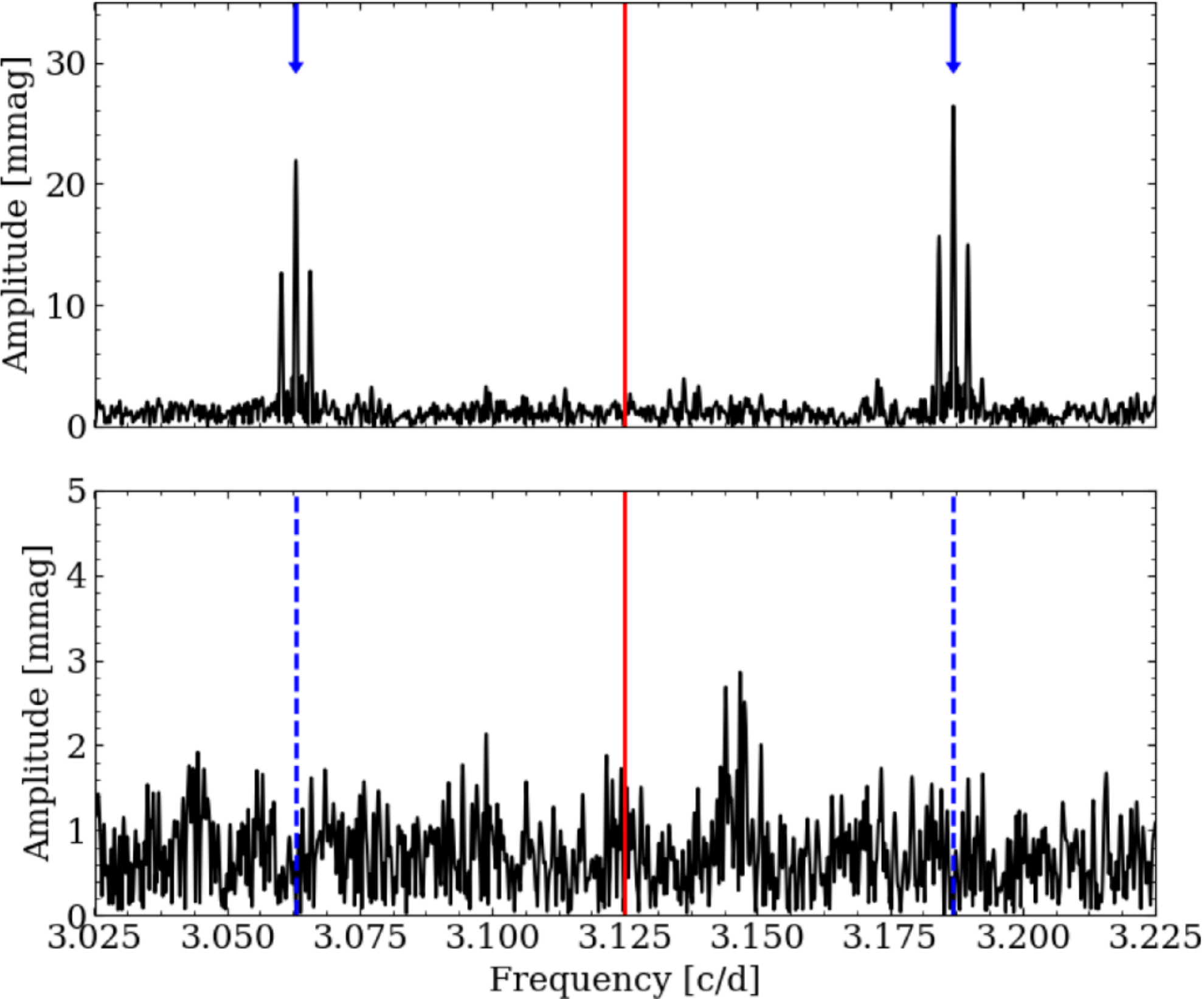}}
\caption{Power spectrum of RRLYR-31383 centered at the first overtone frequency. Top panel shows power spectrum after prewhitening with the first overtone and its harmonics. Bottom panel shows power spectrum prewhitened additionally with presumed Blazhko sidepeaks. Position of the first overtone is marked with red solid line on both panels. Positions of presumed sidepeaks are marked with blue arrows on the top panel and with blue dashed lines on the bottom panel. Blazhko period is $16$\,d.}
\label{fig.example-bl}
\end{figure}

\begin{figure}
\centering
\resizebox{\hsize}{!}{\includegraphics[bb= 0 0 539 478]{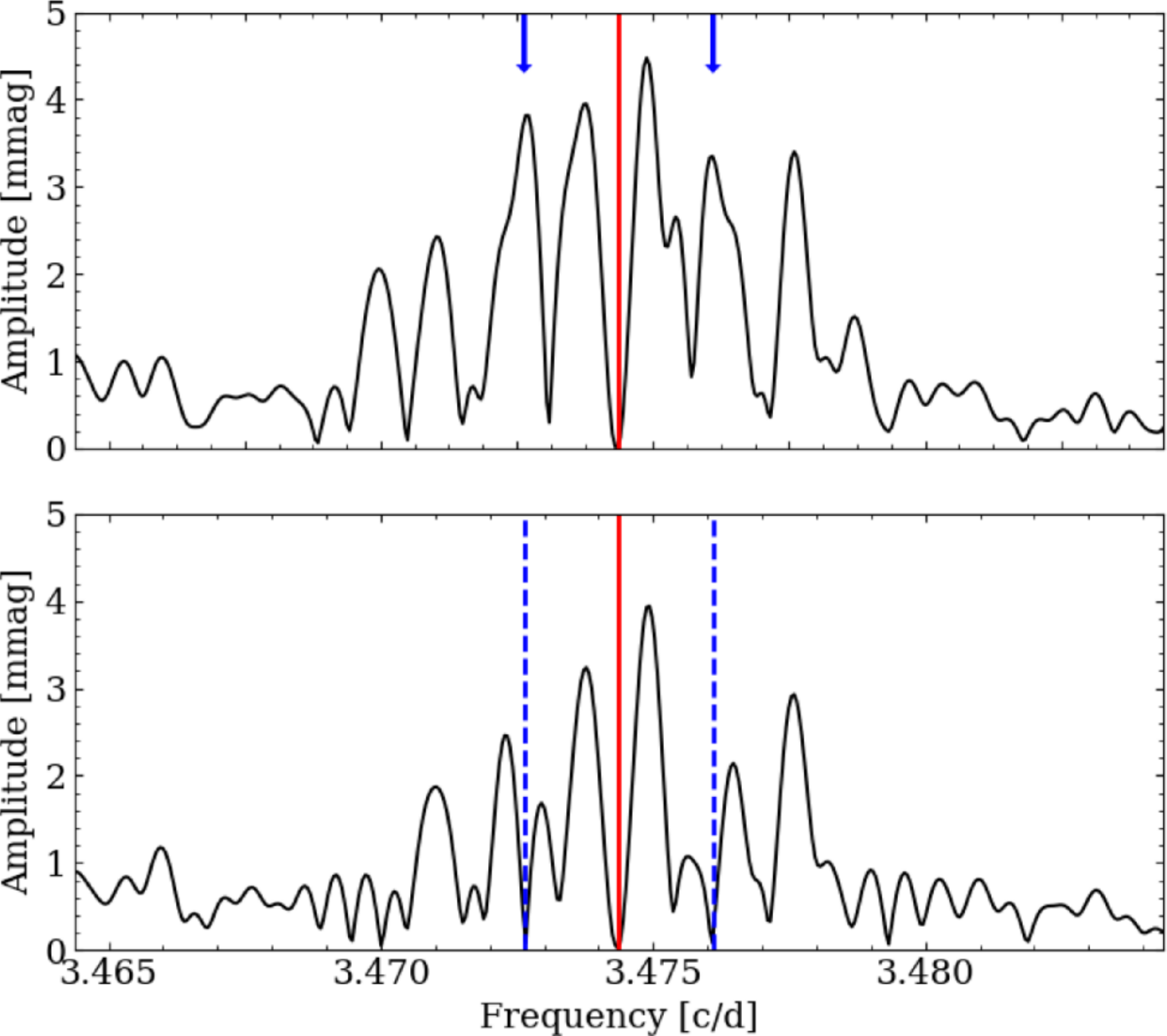}}
\caption{The same as Fig.~\ref{fig.example-bl} but for RRLYR-06746. Period of the presumed Blazhko effect is $588$\,d. This star was not classified as Blazhko star.}
\label{fig.example-niewiem}
\end{figure}

The main problem in the described automatic analysis is due to unresolved remnant power at $f_{\rm 1O}$ after prewhitening with the first overtone. It occurs in 35 percent of RRc stars from our sample. The non-stationary nature of the first overtone frequency is due to irregular changes of pulsation phase and/or amplitude or due to changes on a time-scale too long to be resolved. Time-dependent Fourier analysis shows that in the majority of stars phase (period) changes are dominant.

These unresolved remnant signals in the vicinity of $\fo$, and their unavoidable annual aliases, can be confused with long-period Blazhko effect. Therefore, it was necessary to come up with possibly the best method to distinguish between Blazhko candidates and period changing stars. We decided to inspect visually power spectra of all 3066 candidates before and after prewhitening with the Blazhko components. Examples of inspected plots are in Figs~\ref{fig.example-bl} and \ref{fig.example-niewiem}. Both panels on these figures are centered on a frequency of the first overtone (its position is marked with a red solid line) and show a frequency spectrum after prewhitening with $\fo$ and its harmonics. Top (bottom) panels show the frequency spectrum before (after) prewhitening with the presumed Blazhko components (their positions are marked with arrows in the top panels and with blue dashed lines in the bottom panels). Visual inspection of such plots helped to distinguish between Blazhko effect and non-stationary main period. In Fig.~\ref{fig.example-bl} we present an example of a star with Blazhko modulation. This star does not show period change. Sidepeaks are well resolved with $\fo$ and significant. Star presented in Fig.~\ref{fig.example-niewiem} was classified by automatic procedure as a Blazhko candidate, however inspection of this figure led to exclusion of this star from candidates. This star shows significant period change, which results in an unresolved signal after prewhitening with the first overtone and its harmonics. Annual aliases of this signal are also present. As a result, in the frequency spectrum we observe a broad power excess centred at $\fo$. Automatic procedure classified two peaks within this power excess as Blazhko sidepeaks. These presumed sidepeaks are indicated with arrows on the top panel of Fig.~\ref{fig.example-niewiem}. Note that procedure ignored two signals closest to $\fo$, as they are located within $\fo\pm2/T$, and are unresolved with $\fo$. After prewhitening with the presumed sidepeaks (bottom panel of Fig.~\ref{fig.example-niewiem}) power excess at $\fo$ remains. The star is not classified as Blazhko. Visual inspection of similar figures for all candidates allowed to distinguish between Blazhko effect and period change.

\begin{figure}
\centering
\resizebox{\hsize}{!}{\includegraphics[bb=0 0 431 528]{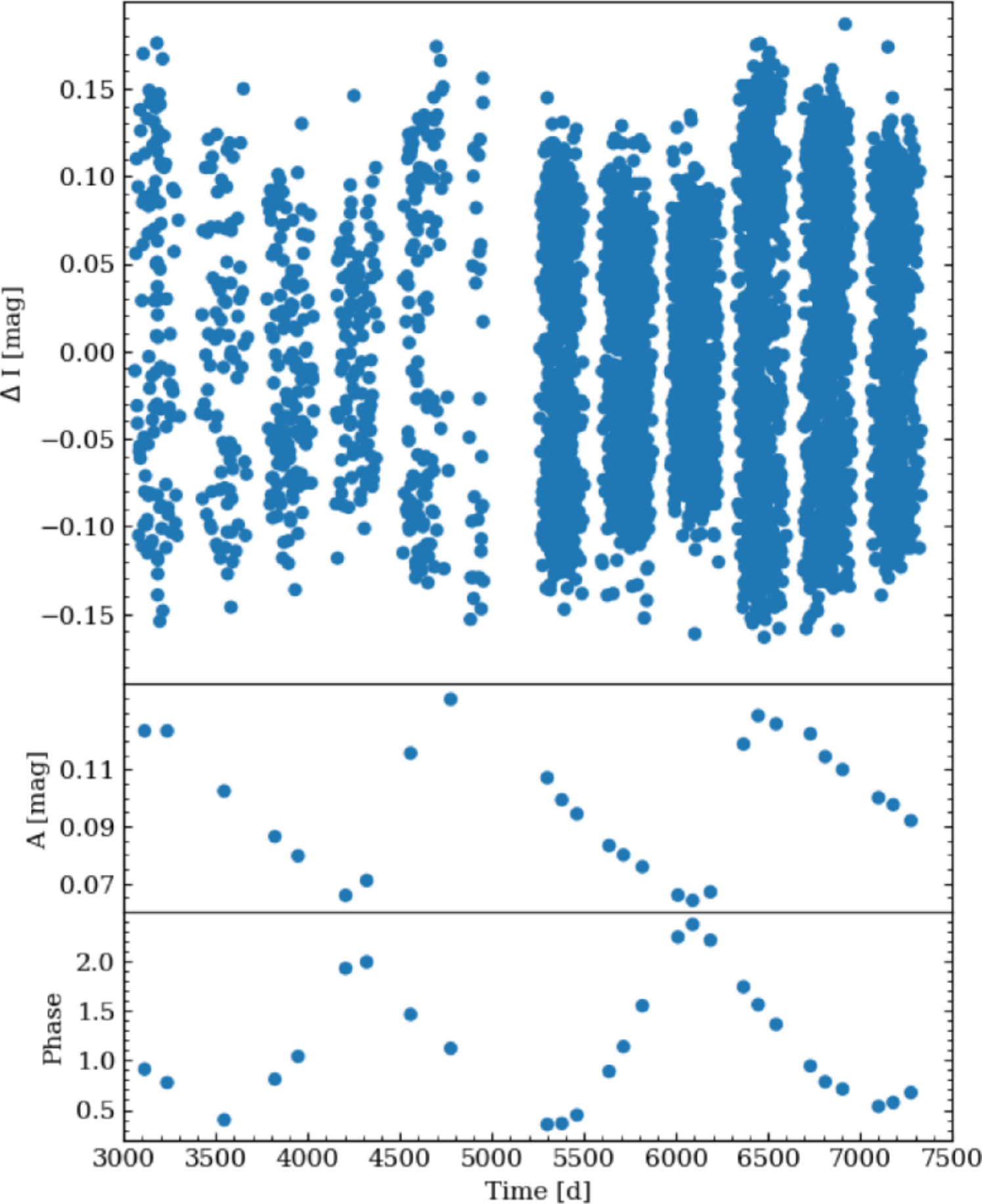}}
\caption{Combined OGLE-III and OGLE-IV data for RRLYR-12659 (top panel). Modulation period is 1769\,d. Middle panel shows changes of amplitude of the first overtone, bottom panel shows changes of its phase.}
\label{fig.12659}
\end{figure}

Candidates which passed the visual inspection were later analyzed manually. Still, there are dubious cases in which the two effects may coexist and are difficult to disentangle. We decided to exclude such cases from the Blazhko sample.

Unresolved power at $\fo$ may also be due to Blazhko effect with period too long to be resolved. To investigate this possibility we merged the OGLE-IV data with data from the previous phases of the OGLE project, if these were available. Longer time base increases the frequency resolution and allows to detect modulation of longer period. As an example, in Fig.~\ref{fig.12659} (top panel) we show data for RRLYR-12659 for which OGLE-III and OGLE-IV data are available. Period of the Blazhko modulation is 1769\,d, and in OGLE-IV data only, the modulation sidepeaks are unresolved. Combining the data allowed to detect long-period modulation. Middle and bottom panels show the results of the time-dependent Fourier analysis \citep{tdfd}. We divided the data into subsets and performed Fourier analysis on each subset separately. As a result, we can study the amplitude changes (middle panel of Fig.~\ref{fig.12659}) and phase changes, which reflect the period change (bottom panel of Fig.~\ref{fig.12659}). Pronounced modulation of both amplitude and phase is clear.

We note that we did not combine data automatically for all stars, because data from previous seasons have typically less dense sampling and hence power spectrum of the combined data has typically a higher noise level. 

\begin{figure*}
 \centering
\begin{minipage}{180mm}
\centering
\resizebox{\hsize}{!}{\includegraphics[bb=0 0 845 650]{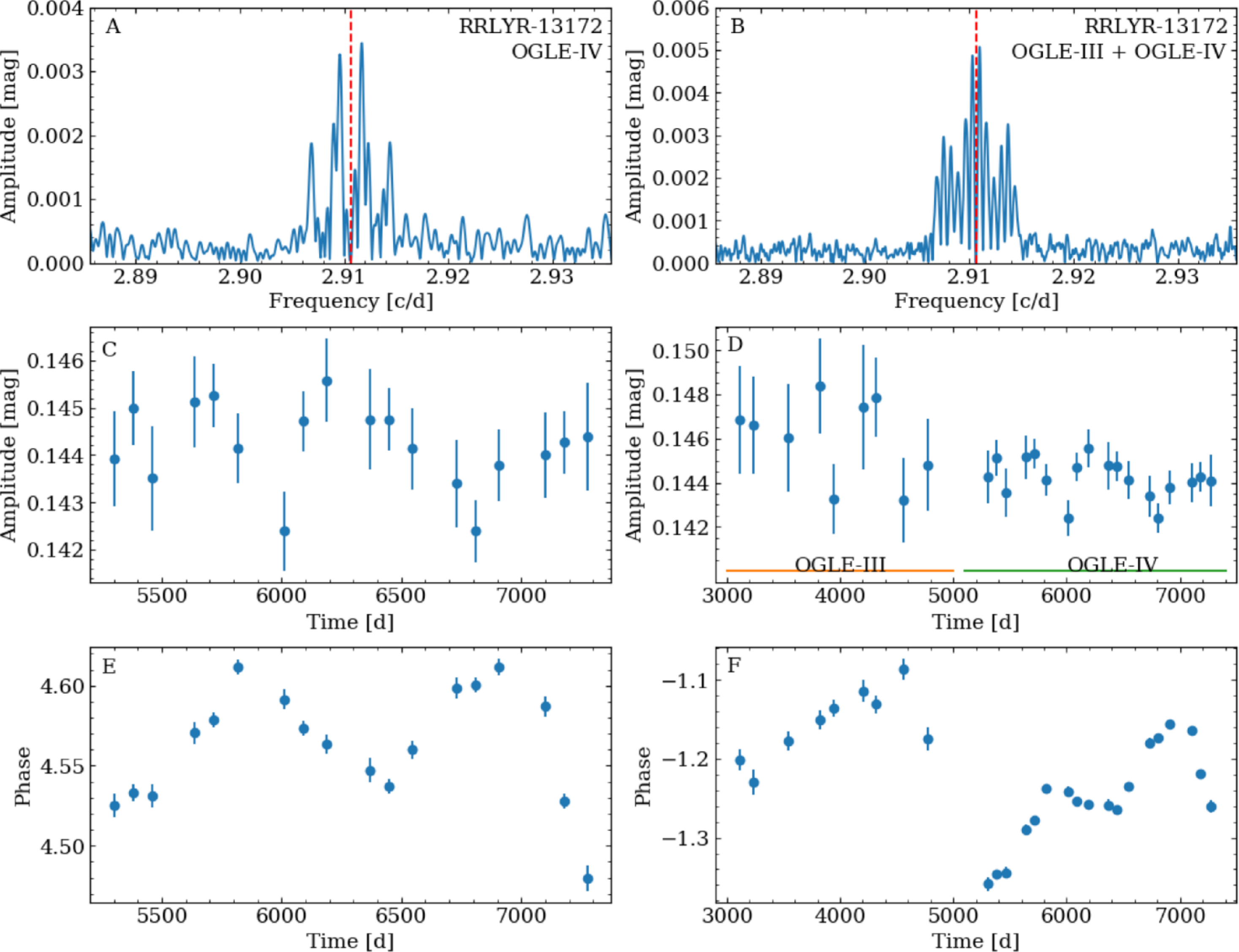}}
\caption{Results of analysis of RRLYR-13172 using OGLE-IV data (left panels) and combined OGLE-III and OGLE-IV data (right panels). Top panels show frequency spectra in the vicinity of the first overtone frequency (marked with red dashed line). Middle and bottom panels show the results of the time-dependent analysis: amplitude (middle panels) and phase (bottom panels) changes.}
\label{fig.13172}
\end{minipage}
\end{figure*}

Combined data were also analyzed for stars with long modulation periods and resolved (within OGLE-IV data) sidepeaks (but close to the adopted resolution criterion). Surprisingly, it resulted in a rejection of some of stars from the Blazhko sample. Example of such star is presented in Fig.~\ref{fig.13172}. Left panels correspond to the analysis of OGLE-IV data, while right panels correspond to the analysis of the combined OGLE-III and OGLE-IV data. Top left panel (A; OGLE-IV) shows frequency spectrum after prewhitening with the first overtone and its eight harmonics. Position of the first overtone is marked with red dashed line. Blazhko sidepeaks are clearly visible, resolved with $\fo$, and accompanied by theirs annual aliases. Corresponding period of modulation is around $950$\,d, so very close to the upper limit of modulation periods that can be resolved with 6 seasons of OGLE-IV data.  Panels C and E show the result of time-dependent analysis: amplitude (panel C) and phase (panel E) change. In the latter panel, quasi-periodic modulation of phase on the same time scale as inferred from the frequency spectrum ($\approx\!950$\,d) is well visible. For amplitude changes (panel C) the errorbars are larger, but modulation is also apparent, in anti-phase to phase modulation. Therefore, based on OGLE-IV data, we would classify this star as Blazhko.

The top right panel of Fig.~\ref{fig.13172} (panel B) shows frequency spectrum of the combined OGLE-III and OGLE-IV data. We observe many signals close to the position of the first overtone, and these are unresolved with the first overtone frequency. Based on this frequency spectrum we would classify this star as a period-changing star. In panels D and F we present the results of the time-dependent analysis applied to the combined data. As opposed to OGLE-IV data only, changes are no longer quasi-periodic, which is best visible for the phase plot. Based on the combined data, the star is rejected from the Blazhko sample.

In general, period changes are common among RRc stars. For some observing seasons, these changes might resemble quasi-periodic modulation. Unfortunately, for some stars with long Blazhko periods there are no earlier data available and the above outlined analysis cannot be conducted. Such stars remain in the Blazhko sample as candidates, as there are no objective reasons to reject them. Admittedly, in some cases we might get fooled.

 \section{Results}\label{sec.results}
 \subsection{Overview}

\begin{figure}
\centering
\resizebox{\hsize}{!}{\includegraphics[bb= 0 0 355 536]{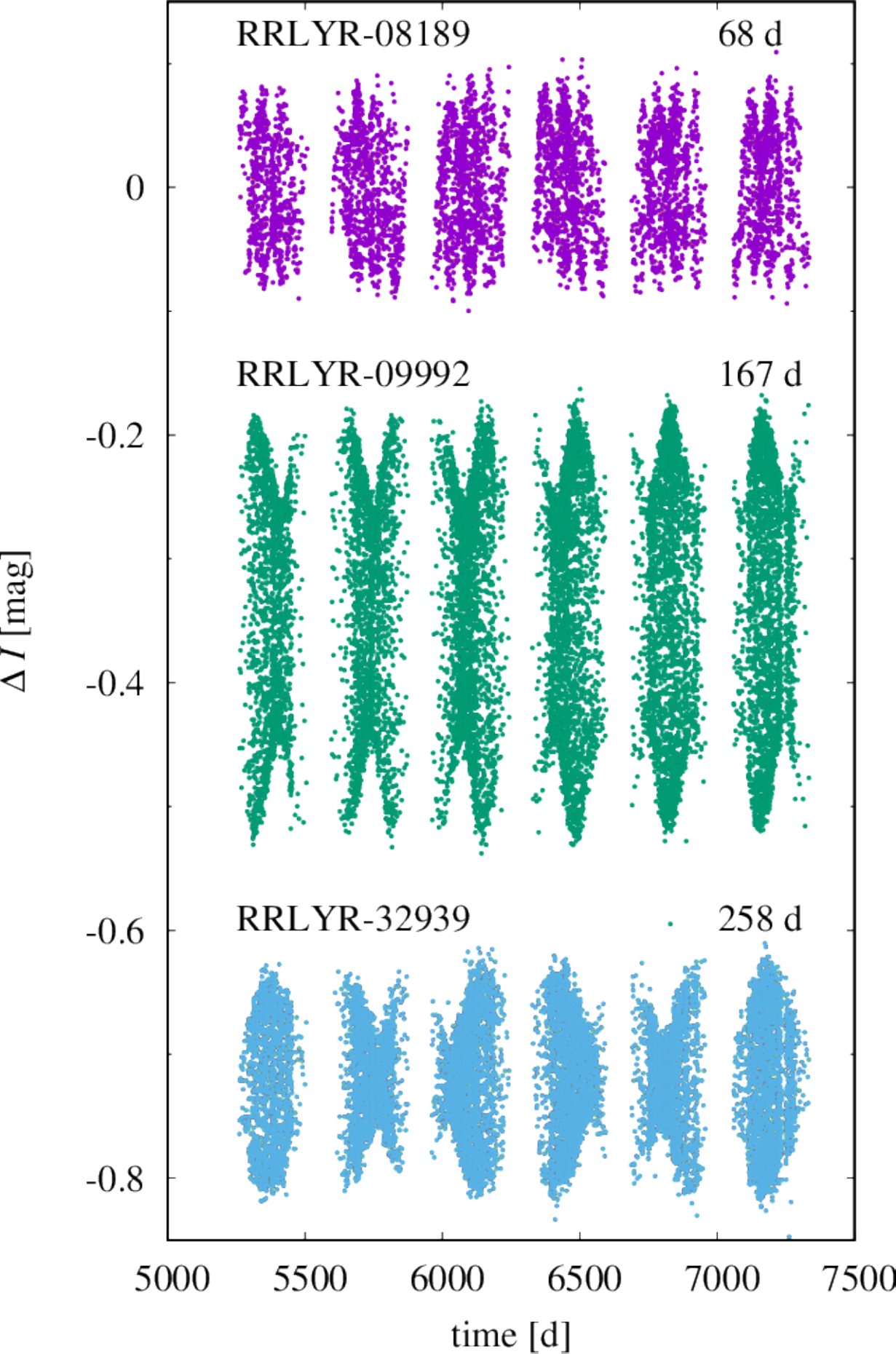}}
\caption{Light curves of Blazhko stars with well visible modulation. Stars are sorted with increasing period of modulation. Star ID is given on the top left side of the light curve. Blazhko period is given on the top right side.}
\label{fig.lc}
\end{figure}

We analysed $10\,826$ first overtone RR~Lyrae stars from the Galactic bulge. After automatic analysis and visual inspection, we have found 607 stars showing evidence of the Blazhko modulation, i.e. either triplets and multiplets (BLm) or dublets (BL1) in their power spectra. They constitute $5.6$\,per cent of the analysed sample. We classified 463 stars to be BLm (76\,per cent of the Blazhko stars) and 144 stars to be BL1 (24\,per cent of the Blazhko stars). Basic parameters for these stars are presented in the Appendix in Tab.~A1, sample of which is presented in Tab.~\ref{tab.sample}. Consecutive columns give: OGLE id, first overtone period, Blazhko type (BLm or BL1), Blazhko modulation period, amplitude of the first overtone, amplitudes of sidepeaks on the high- and low-frequency side of the first overtone, $A_{+}$ and $A_{-}$, respectively, the asymmetry parameter, $Q$, defined as \citep{alcockrrab}:
$$Q=\frac{A_+-A_-}{A_++A_-},$$
and remarks. Some stars have more than one row in the table. In these stars we detected more than one period of modulation. In 175 stars we detected peak at the modulation frequency, $\fmod$, and these stars are marked with `a' in the remarks column of Tab.~A1. In most of the BLm stars, the modulation sidepeaks form a triplet, i.e. their frequencies are $\fo\pm\fmod$. In 85 stars we detected higher order multiplet components (e.g. complete or incomplete quintuplets, or even septuplets). These stars are marked with `b' in the remarks column of Tab.~A1. 

In Fig.~\ref{fig.lc} we present data for three stars in which modulation is visible at the first glimpse, sorted with increasing period of Blazhko modulation. IDs of stars and their Blazhko periods are provided above each light curve. 

\begin{table*}
\centering
\begin{minipage}{170mm}
\caption{Parameters of the Blazhko RRc stars. Consecutive columns provide star's ID, type (BLm or BL1), first overtone and Blazhko periods, amplitudes of the first overtone and the sidepeaks on the higher or lower frequency side, $A_{\rm 1O}$, $A_+$ and $A_-$, respectively, the asymmetry parameter, $Q$, and remarks.}
\label{tab.sample}
\centering
\begin{tabular}{@{}lccccccrl@{}}
\hline
\multicolumn{9}{l}{a - $f_{\rm mod}$ detected in the power spectrum; b - stars with multiplet structures in the power spectrum;}\\ 
\multicolumn{9}{l}{c - stars with additional significant signal; d - multiperiodic Blazhko effect; e - subharmonic of the Blazhko sidepeaks} \\ 
\hline
\hline
 ID & type &$P_{\rm 1O}$\thinspace[d] & $P_{\rm BL}$\thinspace[d] &  $A_{\rm 1O}$\thinspace[mag] & $A_+$\thinspace[mag] &$A_-$\thinspace[mag] & $Q$ & remarks \\
\hline
OGLE-BLG-RRLYR-00065 & BLm & 0.374617(2) & 247(17) & 0.117(3) & 0.025(3) & 0.021(3) & 0.08 &  \\ 
OGLE-BLG-RRLYR-00156 & BLm & 0.379005(2) & 40.15(7) & 0.116(2) & 0.014(2) & 0.021(2) & $-$0.19 &  \\ 
OGLE-BLG-RRLYR-00198 & BLm & 0.3179268(2) & 24.634(2) & 0.0926(7) & 0.0185(7) & 0.0415(7) & $-$0.38 & a \\ 
OGLE-BLG-RRLYR-00416 & BLm & 0.3199298(1) & 2322(65) & 0.0863(6) & 0.0059(7) & 0.0059(7) & 0.00 &  \\ 
OGLE-BLG-RRLYR-00481 & BLm & 0.2652629(1) & 11.500(1) & 0.1127(8) & 0.0246(8) & - & - & a,b \\ 
OGLE-BLG-RRLYR-00569 & BL1 & 0.2725534(7) & 44.61(9) & 0.118(2) & - & 0.028(2) & - &  \\ 
\hline

\end{tabular}
\end{minipage}
\end{table*}

In Tab.~B1 in the Appendix we provide a list of 105 candidates for Blazhko stars. Majority of these stars come from the analysis of combined OGLE data. They can be either Blazhko stars or stars with complex period changes that may mimic quasi-periodic modulation (see the last paragraphs of Sec.~\ref{sec.analysis}). Structure of this table is the same as of Tab.~A1. The incidence rate of Blazhko stars, including the candidates, increases to 6.6 per cent. However, in further analysis, we take into account only stars firmly classified as Blazhko variables, i.e. those from Tab.~A1. The candidates are worth revisiting, once more data are available.

\subsection{Periods of modulation}

\begin{figure}
\centering
\resizebox{\hsize}{!}{\includegraphics[bb= 0 0 406 308]{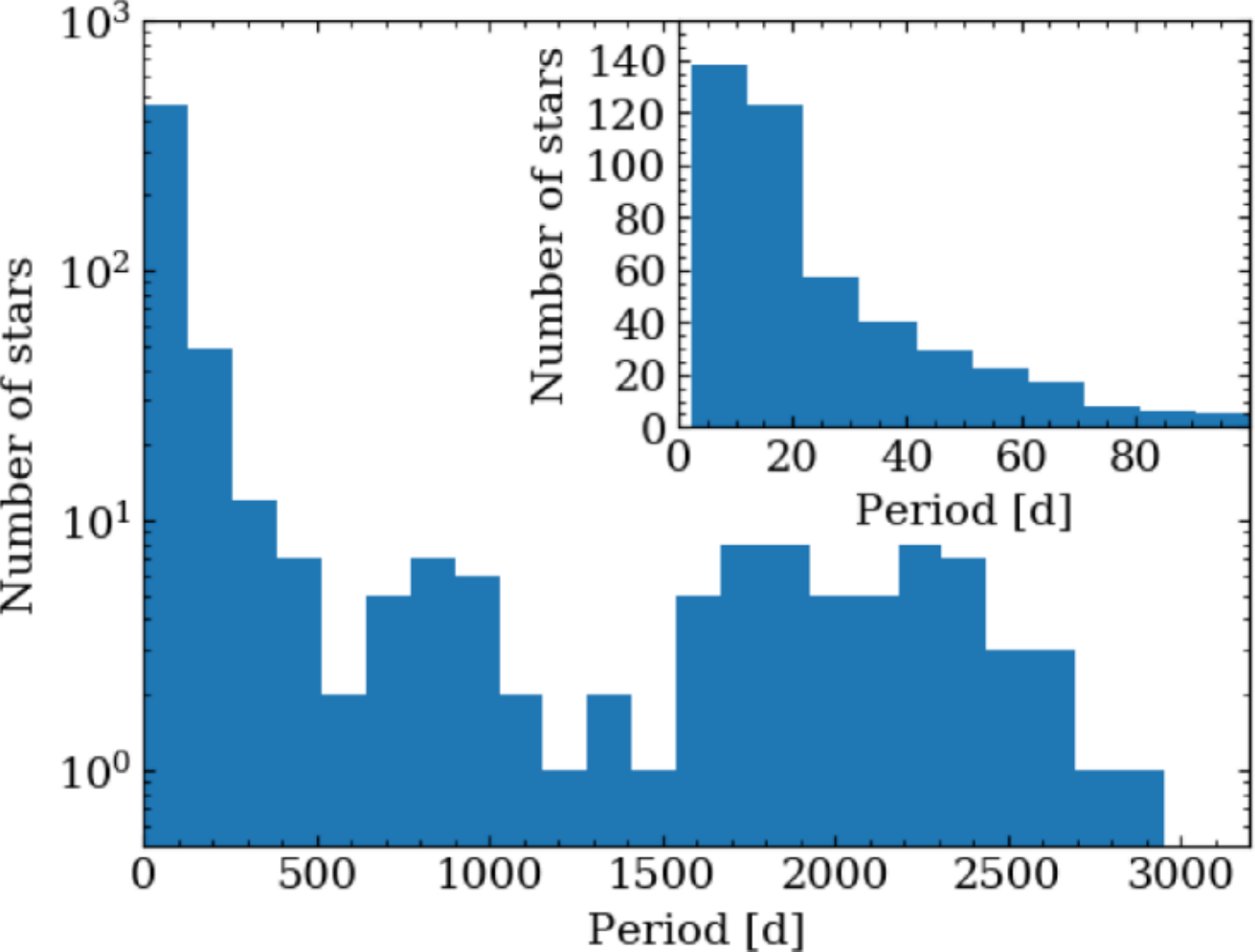}}
\caption{Histogram of Blazhko periods for RRc stars. In top right corner of the plot we include zoom for short Blazhko period range. Note that number of stars in the zoom is plotted in linear scale, whereas the whole distribution is plotted in logarithmic scale.}
\label{fig.pbl}
\end{figure}

In Fig.~\ref{fig.pbl} we present the distribution of Blazhko periods in our sample. In the top right corner of this figure we present zoom for the short period part of the distribution (periods lower than 100\,d). Note that in the zoom, the vertical axis is in linear scale, whereas the whole distribution is plotted in logarithmic scale. It is easily visible, that majority of Blazhko stars have rather short periods of modulation. Number of stars is the largest for modulation periods below 20\,d and quickly decreases with increasing modulation period. Median value of the Blazhko period in our sample is 29\,d, whereas average period is 273\,d. The shortest detected period is 2.23\,d in RRLYR-24030 (see Sect.~\ref{ssec:24030}), which is shorter than the shortest Blazhko period for an RRc star known so far, which is above 5\,d \citep{skarka}. In our sample, there are 13 stars with Blazhko period shorter than 5\,d. Interestingly, only four of them are classified as BLm.

In Fig.~\ref{fig.pbl} we observe a slight decrease of stars for Blazhko periods above 500\,d. It is most likely caused by detection problems. In power spectra of stars with presumed Blazhko periods around 500\,d and above, Blazhko sidepeaks are mutual annual aliases. In most of these stars, Blazhko effect is uncertain, and consequently, they were rejected from the Blazhko sample.

The longest possible Blazhko period depends on the time base of observations. For the majority of stars we used 6 seasons of OGLE-IV data. This data length, on average, allows us to detect modulations with periods up to 1000\,d. Hence in Fig.~\ref{fig.pbl} we observe a drop of stars for periods around 1000\,d. Fortunately, for some stars observations from the previous OGLE phases are available. Combining all available OGLE data allowed to find Blazhko stars with longer modulation periods. These stars are included in the distribution in Fig.~\ref{fig.pbl}. There are only a few such stars, therefore the distribution above 1000\,d does not have a smooth shape. The longest detected period in the combined OGLE data is 2954\,d for RRLYR-02478.

\subsection{Multiple Blazhko effect}

\begin{figure}
\centering
\resizebox{\hsize}{!}{\includegraphics[bb= 0 0 406 304]{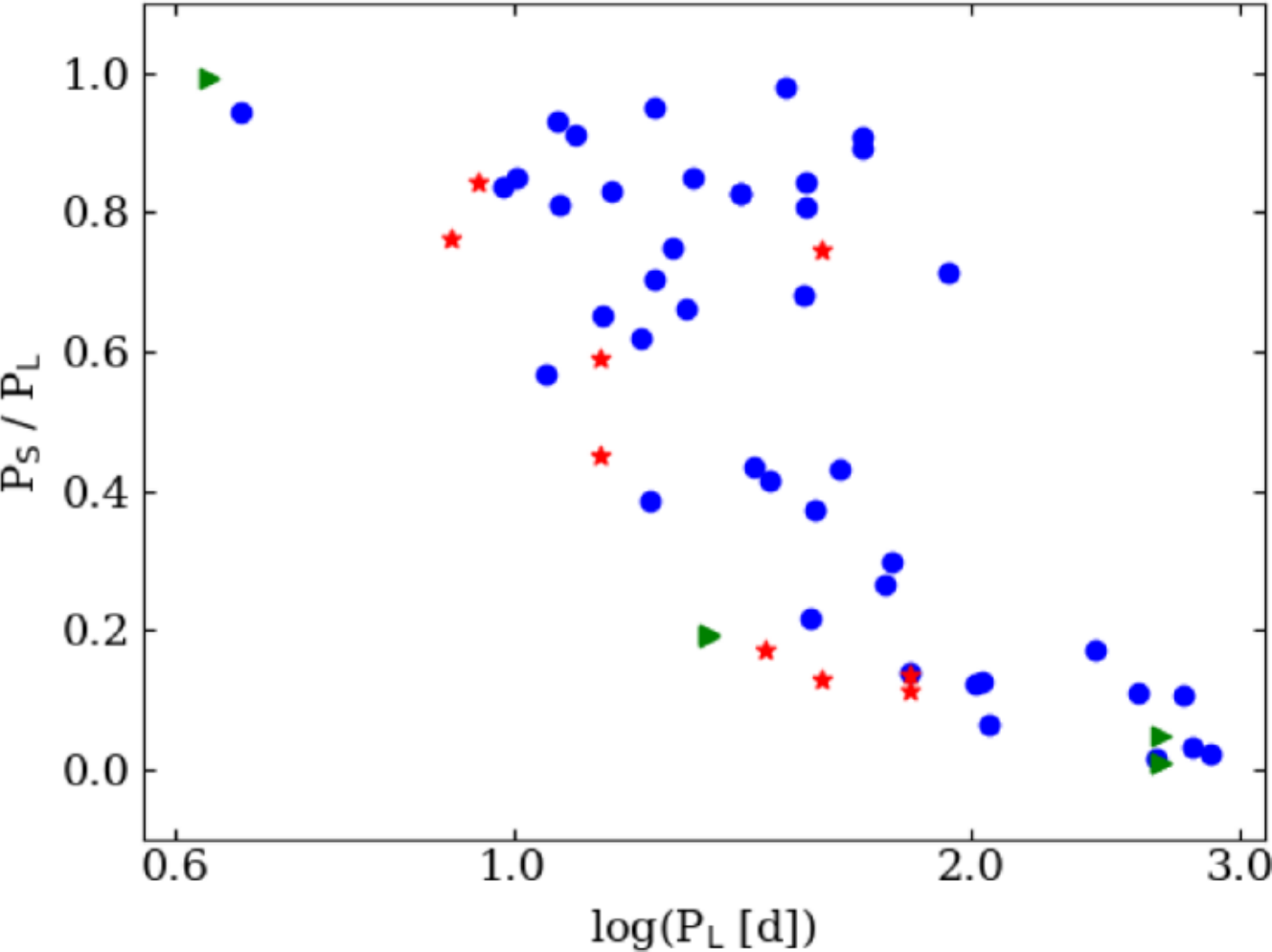}}
\caption{Period ratio diagram for stars with multiple Blazhko periods. Blue points correspond to stars with two Blazhko periods. Red asterisks are for stars with three Blazhko periods. Green triangles are plotted for single star with four Blazhko periods.}
\label{fig.pet_2bl}
\end{figure}

In 47 stars in our sample we detected more than one Blazhko period. These stars have more than one line of data in Tab.~A1. In 43 stars we detected two Blazhko periods, in three stars we found three periods (RRLYR-33726, RRLYR-06207, RRLYR-06625; more detailed analysis in Sect.~\ref{ssec:threemod}), and in one star we observe four different periods (RRLYR-32935; more detailed analysis in Sect.~\ref{ssec:fourmod}). Majority of these stars have rather short modulation periods. For 36 stars all modulation periods are shorter than $50$\,d. Only in five stars we observe significantly longer periods (above $100$\,d). We do not observe any rule whether the primary Blazhko period (associated with higher amplitude of modulation) is shorter or longer than the period of secondary modulation(s). Roughly half of stars with double-periodic Blazhko effect have the primary period shorter than the secondary period.

We illustrate the modulation periods in the period ratio diagram in Fig.~\ref{fig.pet_2bl}. On horizontal axis we plot the longer Blazhko period. On the vertical axis we plot the ratio of the shorter to longer Blazhko period. Stars with two modulation periods are marked with blue points in Fig.~\ref{fig.pet_2bl}, stars with three periods with red asterisks (three symbols per star) and star with four Blazhko periods with green triangles (four symbols). Observed period ratios cover a wide range. We do not observe any particular grouping in the period ratio diagram. For long Blazhko periods we observe low period ratios.

\subsection{Amplitudes of modulation}

\begin{figure}
\centering
\resizebox{\hsize}{!}{\includegraphics[bb= 0 0 576 382]{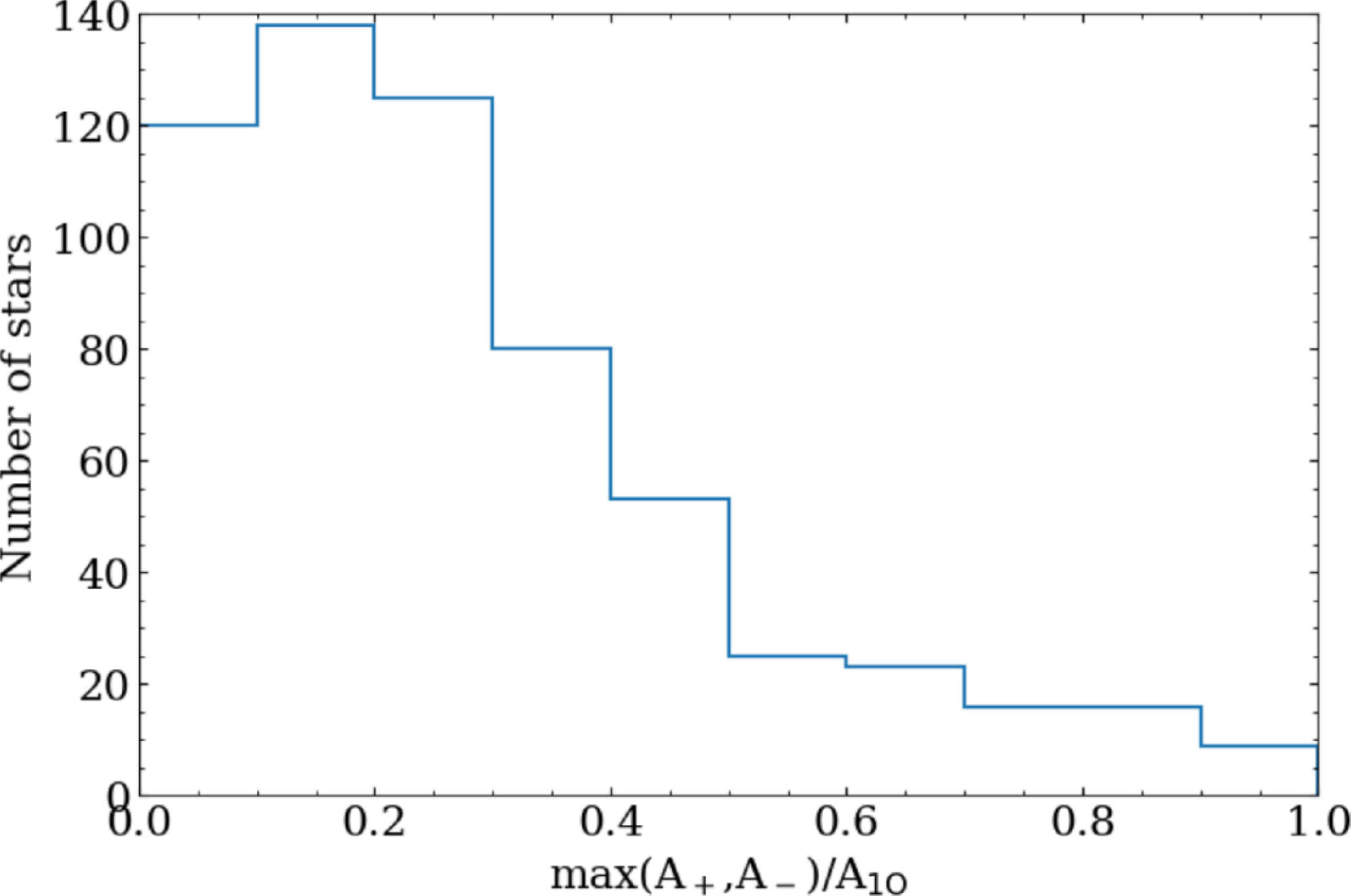}}
\caption{Distribution of relative modulation amplitude for Blazhko RRc stars.}
\label{fig.amp_ratio}
\end{figure}

In Fig.~\ref{fig.amp_ratio} we present a distribution of relative modulation amplitude -- ratio between amplitude of the higher of the modulation sidepeaks detected at the first overtone frequency, and amplitude of the first overtone, $\max(A_+,A_-)/A_{\rm 1O}$. For BL1 stars we take the amplitude of the only detected sidepeak. For 85\,per cent of stars the relative modulation amplitude is below $0.5$. However, for some stars we observe large values; a record holder is RRLYR-03155 (BLm) in which the relative modulation amplitude is close to $1$.

\begin{figure}
\centering
\resizebox{\hsize}{!}{\includegraphics[bb= 0 0 576 377]{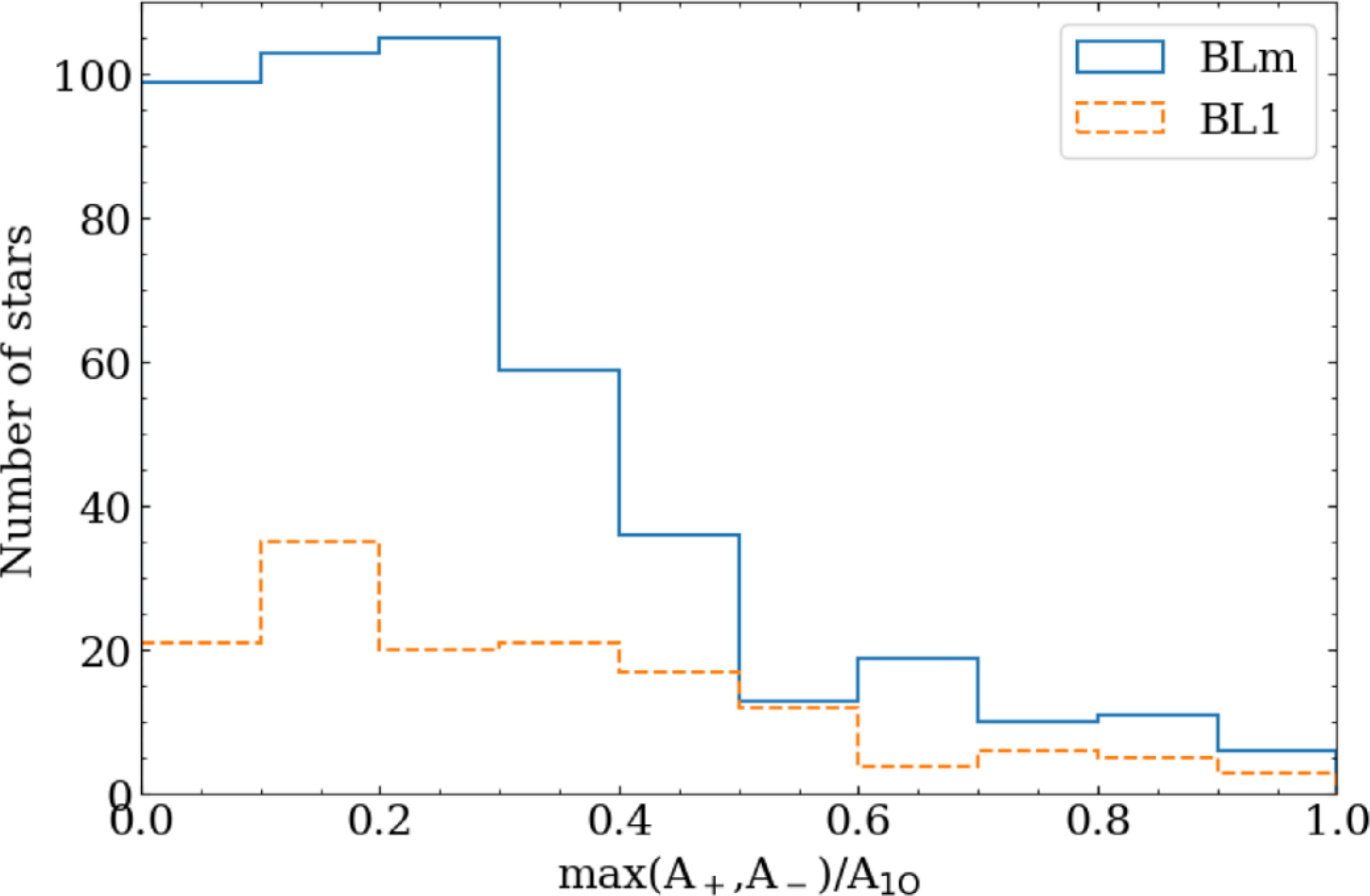}}
\caption{Distribution of relative modulation amplitude with distinction between BLm (solid blue line) and BL1 (dashed orange line) stars.}
\label{fig.amp_ratio2}
\end{figure}

In Fig.~\ref{fig.amp_ratio2} we present a comparison between relative modulation amplitudes for BLm and BL1 stars. In both groups, stars tend to have relative modulation amplitudes below $0.5$. 62 BLm stars (13\,per cent of BLm stars) have relative modulation amplitude above $0.5$. Average value for BLm stars is$0.28\pm0.01$. In the case of BL1 stars, distribution is more flat. Still there are more BL1 stars with relative modulation amplitude below $0.5$, but the difference between number of stars with low (below $0.5$) and high (above $0.5$) relative modulation amplitude is smaller than in the case of a BLm group. 21\,per cent of BL1 stars have relative modulation amplitudes above $0.5$. Average value of relative modulation amplitude for BL1 stars is $0.33\pm0.02$, so a bit larger than for BLm stars.

\begin{figure}
\centering
\resizebox{\hsize}{!}{\includegraphics[bb= 0 0 555 389]{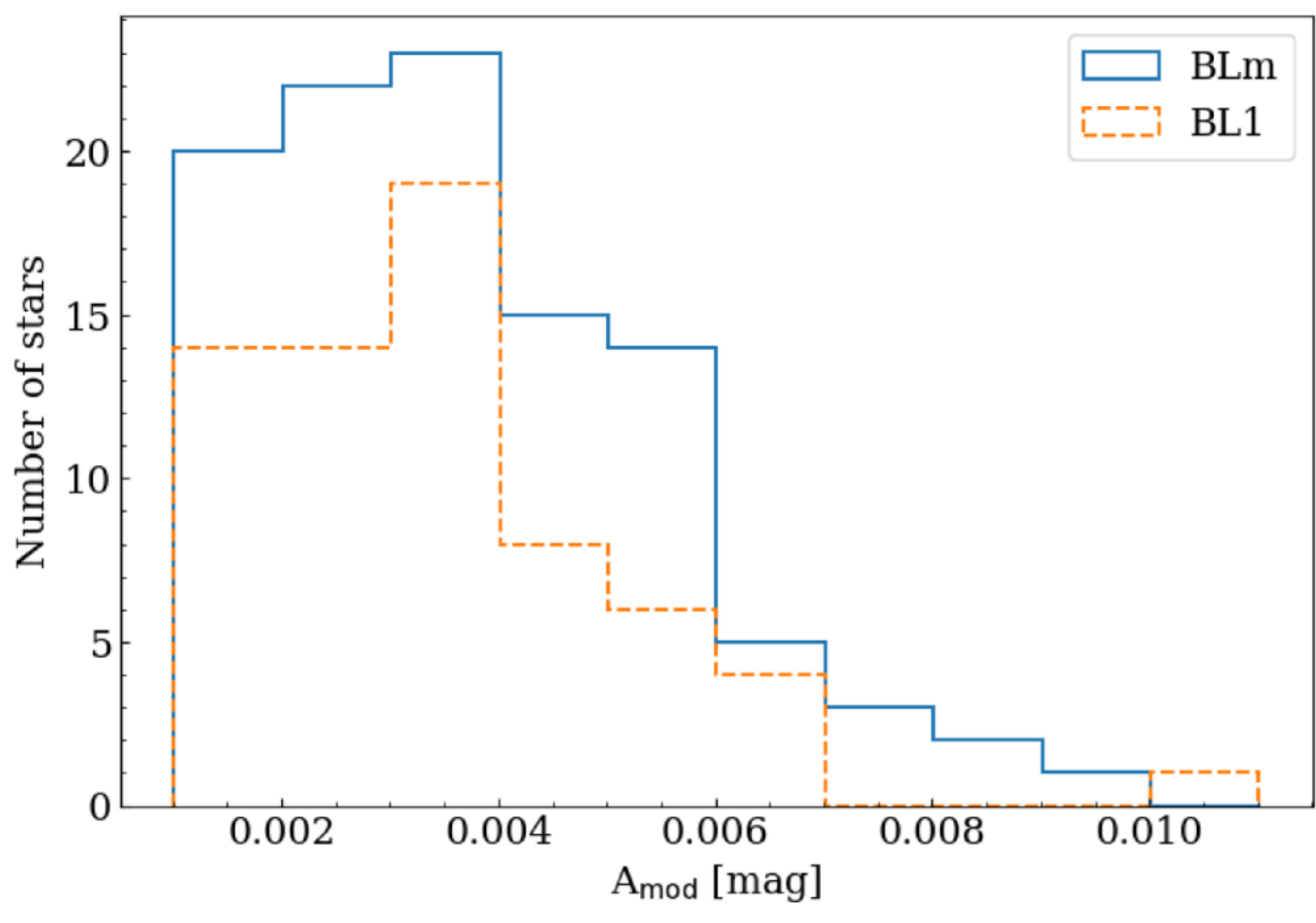}}
\caption{Distribution of mean brightness modulation amplitudes for BLm (blue solid line) and BL1 (orange dashed line) stars.}
\label{fig.a_hist}
\end{figure}

In Fig.~\ref{fig.a_hist} we present a distribution of mean brightness modulation amplitudes, which are amplitudes of signals detected at the modulation frequency, $\fmod = 1/P_{\rm m}$. BLm stars are plotted with solid blue line, while BL1 stars are plotted with orange dashed line. Distributions are similar for the two groups. Signal at the modulation frequency was detected in 66 BL1 stars, which corresponds to 46\, per cent of all BL1 stars. Among BLm stars we found 109 such stars, which constitutes 24\,per cent. Typically, amplitudes of mean brightness modulation are low, in a milimagnitude range. Majority of stars have mean brightness modulation amplitudes below 5\,mmag. Only three stars have amplitude of mean brightness modulation higher than 10\,mmag. Two of them are BL1 stars and one is BLm star. The highest amplitude, 17\,mmag, is detected for BLm star (RRLYR-24030).
 
\begin{figure}
\centering
\resizebox{\hsize}{!}{\includegraphics[bb= 0 0 581 375]{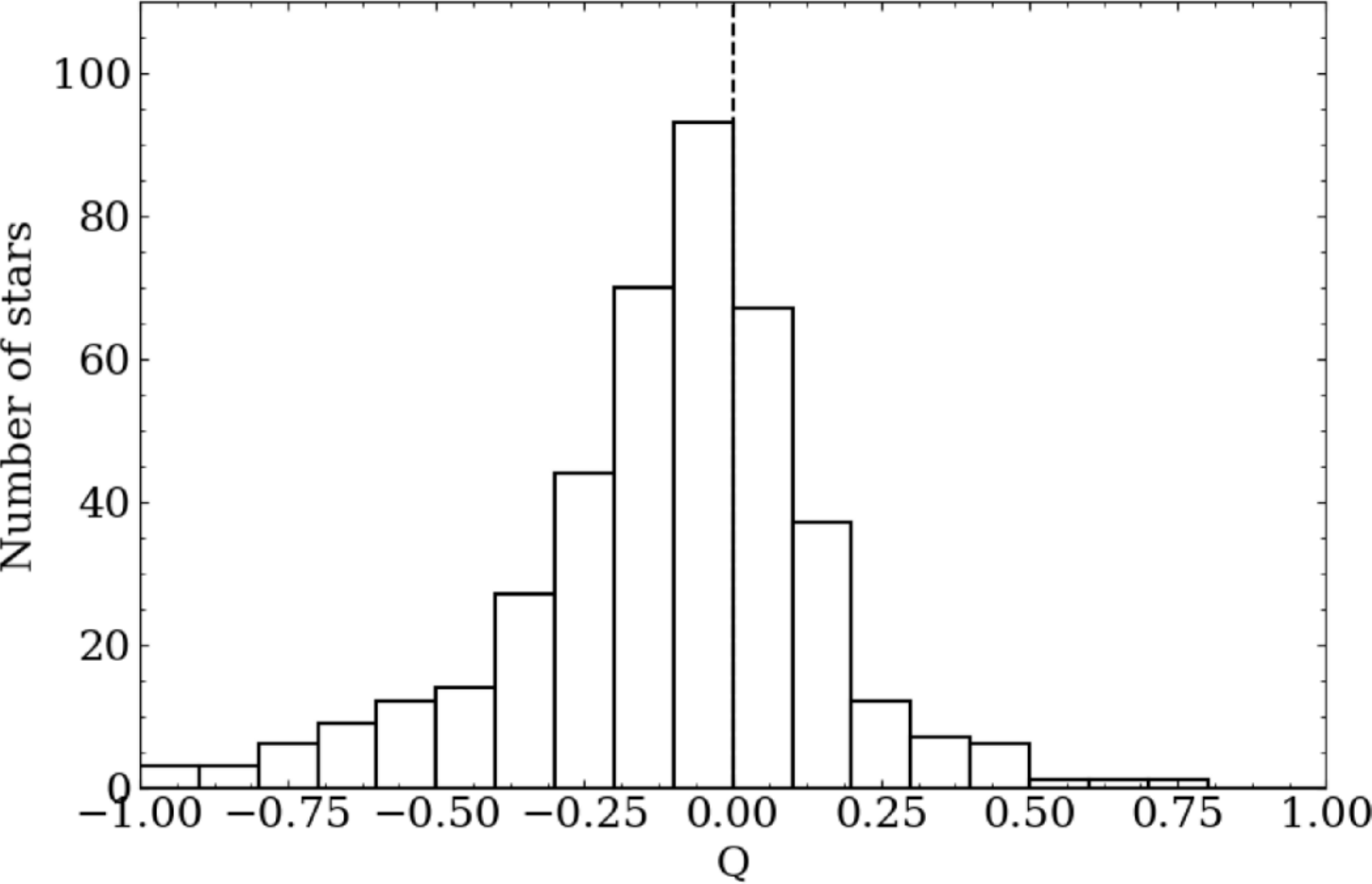}}
\caption{Distribution of $Q$ parameter for BLm stars. Vertical dashed line corresponds to $Q=0$.}
\label{fig.q_param}
\end{figure}

In order to investigate asymmetry in amplitudes of the sidepeaks, we used the asymmetry parameter $Q$. The parameter can be calculated only for stars with sidepeaks present on both sides of the main signal -- for the majority of BLm stars (in some BLm stars, incomplete multiplets were detected, with all sidepeaks on one side of the radial mode frequency). Distribution of the $Q$ parameter is presented in Fig.~\ref{fig.q_param}. The distribution is asymmetric. 281 stars have negative $Q$ (68\,per cent), which means that the low-frequency sidepeaks are dominant in most stars from our sample. The average value of $Q$ parameter is $-0.1$. For BL1 stars and BLm stars with sidepeaks only on one side of the first overtone frequency, we can assign $Q=+1$ for stars with sidepeak on the higher frequency side and $Q=-1$ for the sidepeak on the lower frequency side. Interestingly, we observe no preference towards one of the values. 97 of stars have $Q=+1$ and 98 stars have $Q=-1$. 

\subsection{Additional signals in Blazhko stars}\label{ssec:as}
   
\begin{figure}
\centering
\resizebox{\hsize}{!}{\includegraphics[bb= 0 0 347 280]{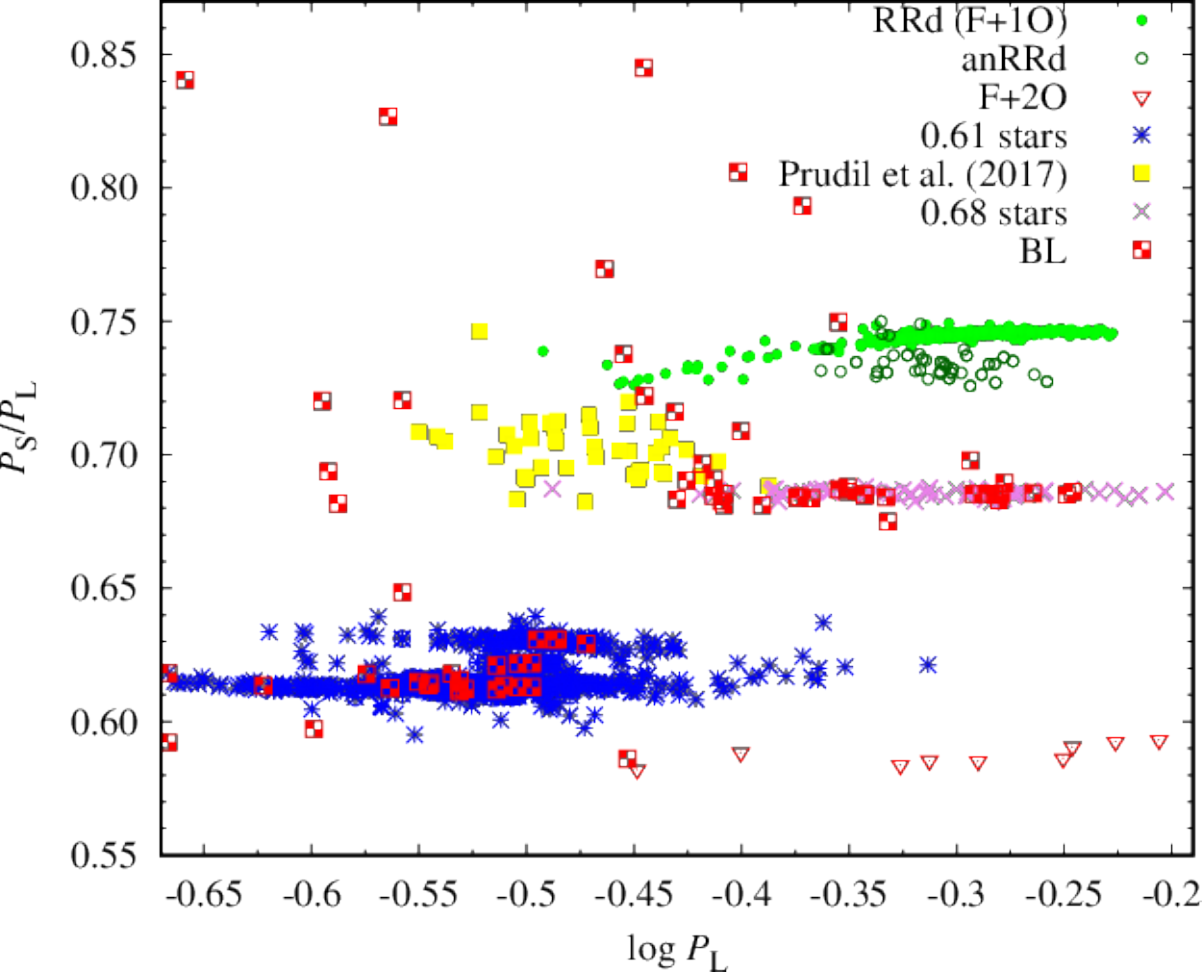}}
\caption{Petersen diagram for multi-periodic RR~Lyrae stars. Different symbols denote different groups of double-periodic stars as indicated in the key and discussed in more detail in the text.}
\label{fig.pet}
\end{figure}

Besides first overtone frequency, its harmonics, and modulation frequencies, we have found additional significant signals in 104 stars. These stars are marked with `c' in the remarks column in Tab.~A1. Full list of stars with additional signals is included in Tab.~C1 in the Appendix, sample of which is presented in Tab.~\ref{tab.add} for a reference. Consecutive columns give star's ID, period of the first overtone, $P_{\rm 1O}$, and of the additional variability, $P_{\rm x}$, ratio of these two periods, $P_{\rm S}/P_{\rm L}$ (shorter to longer), amplitude of the first overtone, $A_{\rm 1O}$, and amplitude ratio, $A_{\rm x}/A_{\rm 1O}$. The last column gives possible explanation of the signal: $RR_{0.61}$, $RR_{0.68}$, or anomalous RRd star (anRRd), as discussed in the next paragraphs. Single signals close to the first overtone frequency, which can be either due to the Blazhko effect or due to non-radial pulsation, are marked with nr/BL. Some stars have more than one row in the table. In these stars we detected more than one additional frequency.

In Fig.~\ref{fig.pet} we present the Petersen diagram (ratio of the shorter to longer pulsation period as a function of the logarithm of the longer period) of already known groups of multi-periodic RR~Lyrae stars. RRd stars are marked with small green points. They follow a tight progression in the Petersen diagram. In the same part of the diagram, anomalous RRd stars are located \citep[][dark green circles in Fig.~\ref{fig.pet}]{soszrrd}. As compared to classical RRd stars, their period ratios are anomalous, typically lower than in the classical RRd stars of the same period. In addition, in anomalous RRd stars, it is the fundamental mode that typically dominates the pulsation, in contrast to classical RRd stars. Most interestingly, majority of these stars show Blazhko modulation. These stars were found in the Galactic bulge \citep{rrdbl}, Magellanic Clouds \citep{soszrrd} and in the globular clusters, M3 \citep{jurcsik3} and NGC\,6362 \citep{ngc}.

Red triangles in Fig.~\ref{fig.pet}, mark stars pulsating in the fundamental mode and the second overtone simultaneously. Majority of these stars show the Blazhko effect \citep{benko2}.

Interesting group of double-mode stars is marked with blue asterisks. The main pulsation mode in these stars is the first overtone. In addition, a shorter-period variability is detected in these stars, forming characteristic period ratio with the first overtone period, between $0.60$ and $0.64$. For the first time, this additional signal was detected in an RRd star AQ~Leo \citep{aqleo}. Later studies resulted in more such stars belonging to different stellar systems, observed both from ground and from space \citep[e.g.][]{lmc, 068kepler}. However, the majority of these stars were discovered in the OGLE-III and OGLE-IV data from the Galactic bulge \citep{061o3, 061o4}\footnote{\label{note1}Full analysis of all Galactic bulge RRc stars, resulting in a significant number of new detections of $RR_{0.61}$ and $RR_{0.68}$ stars, will be published soon (Netzel et al., in prep). Fig.~\ref{fig.pet} includes newly discovered stars.}. At present, this group of stars is considered to be separate group of double-mode radial--non-radial stars, denoted $RR_{0.61}$. \cite{dziembowski} proposed that the additional periodicities in these stars correspond to harmonics of non-radial modes of moderate degrees.

Pink crosses in Fig.~\ref{fig.pet}, mark another group of double-periodic stars, $RR_{0.68}$. In these stars the main pulsation mode is the first overtone and additionally we observe low-amplitude signal with longer period. Nineteen such stars were identified in the OGLE-IV data \citep{068, 068pta}$^{\rm 2}$ and one star was identified in the {\it Kepler} data \citep{068kepler}. Explanation of this group has not been found yet \citep[see however][]{dziembowski}.

Yellow squares mark double-periodic stars detected in the OGLE Galactic bulge data by \cite{rrlyr}. 38 of these stars were previously classified as RRab and 4 as RRc. Additional periodicity is always of shorter period than the dominant periodicity and has relatively large amplitude (typically 20\,per cent of the main pulsation mode amplitude). In 24\,per cent of these stars Blazhko modulation was detected.

RRc stars with the Blazhko effect and additional signals detected in the frequency spectrum are marked on top of the above discussed groups with red symbols. Some Blazhko stars fit very well to the already known groups. One star falls in the Petersen diagram very close to the sequence of RRd stars. This star might be anomalous RRd star. However, it would be a somewhat atypical member of this group, as first overtone dominates its pulsation and period ratio is a bit higher than in the classical RRd stars. Similar cases are known among anomalous RRd stars, but are very rare. 19 stars fit very well within the sequence defined by $RR_{0.61}$ stars. 24 stars fall within the sequence formed by $RR_{0.68}$ stars. In 48 stars we observe very high period ratio (above 0.9). In these stars we detected additional signal with a frequency very close to the frequency of the first overtone. This signal can be either a non-radial mode with a frequency close to the frequency of the main pulsation mode, or it can be due to the Blazhko effect, but since we did not detect other sidepeaks we did not classified these stars as Blazhko. 

Detailed discussion of additional signals detected in all RRc stars from the OGLE Galactic bulge collection, including Blazhko stars reported here, will be included in the dedicated paper, Netzel et al., in prep.

 \begin{table*}
  \begin{minipage}{150mm}
   \caption{Stars with additional frequencies in power spectrum. Subsequent columns contain star's ID, first overtone period, additional period, period ratio, first overtone amplitude and amplitude ratio. Last column contains the possible explanation of the additional signal: $RR_{0.61}$, $RR_{0.68}$, anRRd or nr/BL.}
   \label{tab.add}
   \begin{tabular}{@{}||lcccccl@{}}
 
 ID& $P_{\rm 1O}$ [d] & $P_{\rm X}$ [d] &$P_S/P_L$ &${A_{\rm 1O}}$ [mag] &$A_{\rm 1O}/A_{\rm X}$& Possible explanation\\
  \hline
OGLE-BLG-RRLYR-00719 & 0.26569 & 0.26319 & 0.99059 & 0.11564 & 0.0605 & nr/BL \\
OGLE-BLG-RRLYR-00959 & 0.27026 & 0.26688 & 0.98747 & 0.08773 & 0.1281 & nr/BL \\
 & 0.27026 & 0.56903 & 0.47496 & 0.08773 & 0.0527 &  \\
 & 0.27026 & 0.13918 & 0.51497 & 0.08773 & 0.0567 &  \\
OGLE-BLG-RRLYR-01283 & 0.31583 & 0.31109 & 0.98497 & 0.10217 & 0.0478 & nr/BL \\
 \end{tabular}
 \end{minipage}
 \end{table*}

\section{Comparison of properties of Blazhko and non-Blazhko stars}\label{sec:blvsnbl}

In this section we compare the global properties of Blazhko and non-Blazhko RRc stars from the Galactic bulge. Properties which are considered are mean brightness, period and amplitude of pulsations, and light curve shape.

\begin{figure}
\centering
\resizebox{\hsize}{!}{\includegraphics{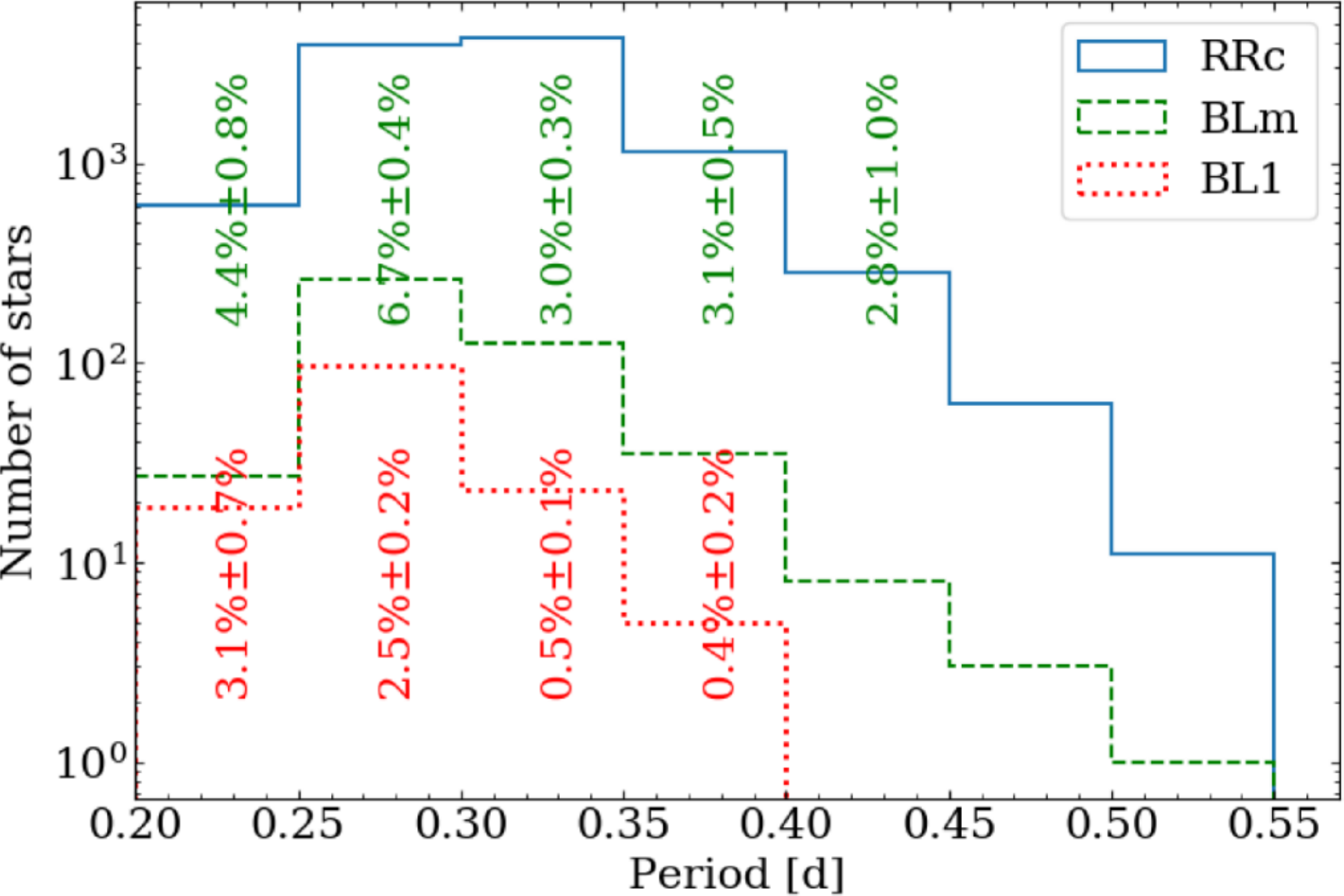}}
\caption{Distribution of periods of the radial first overtone for all RRc stars and for Blazhko RRc stars based on OGLE-IV data. Distribution for all RRc stars is plotted with blue solid line. BLm stars are plotted with green dashed line and BL1 stars are plotted with red dotted line. Incidence rates with statistical errors are given in the plot. The top row corresponds to BLm incidence rates, the bottom row corresponds to BL1 incidence rates. Statistical errors are calculated with an assumption that the population follows the Poisson distribution \citep[e.g.][]{alcockrrab}. Incidence rate is given only if number of stars in the bin is bigger than five.}
\label{fig.p1o}
\end{figure}

In Fig.~\ref{fig.p1o} we plot the distribution of first overtone periods for all RRc stars (blue solid line) and for Blazhko stars: BLm stars (green dashed line) and BL1 stars (red dotted line). In the plot we provided the incidence rates of BLm (top row) and BL1 (bottom row) stars in each bin, together with their errors. Blazhko stars have shorter periods of pulsation than non-modulated stars. The highest incidence rate for BLm stars, 6.7\,per cent, is in the bin with pulsation periods in the $0.25-0.3$\,d range. BLm stars are hardly present among stars with pulsation periods above $0.45$\,d. BL1 stars tend to have even shorter periods. The highest incidence rate is in the bin with the shortest pulsation periods, i.e. $0.2-0.25$\,d. The incidence rate for BL1 stars quickly decreases for bins with longer periods. BL1 stars are very scarce for pulsation periods longer than $0.3$\,d.

Peak-to-peak amplitudes of pulsations of BLm and BL1 stars cover similar range as non-Blazhko RRc stars and, in general, follow the same progression. An average pulsation amplitude for BLm stars is 0.23$\pm$0.05\,mag, whereas for BL1 stars average amplitude is 0.21$\pm$0.06\,mag, so average amplitudes are the same within errors. One significant outlier is RRLYR-24030, BLm star with the highest amplitude in the sample (0.5214\,mag). Also, this star has the shortest Blazhko period (2.23\,d). This star is discussed in more detail in Sec.~\ref{ssec:24030}.

\begin{figure}
\centering
\resizebox{\hsize}{!}{\includegraphics[bb= 0 0 570 377]{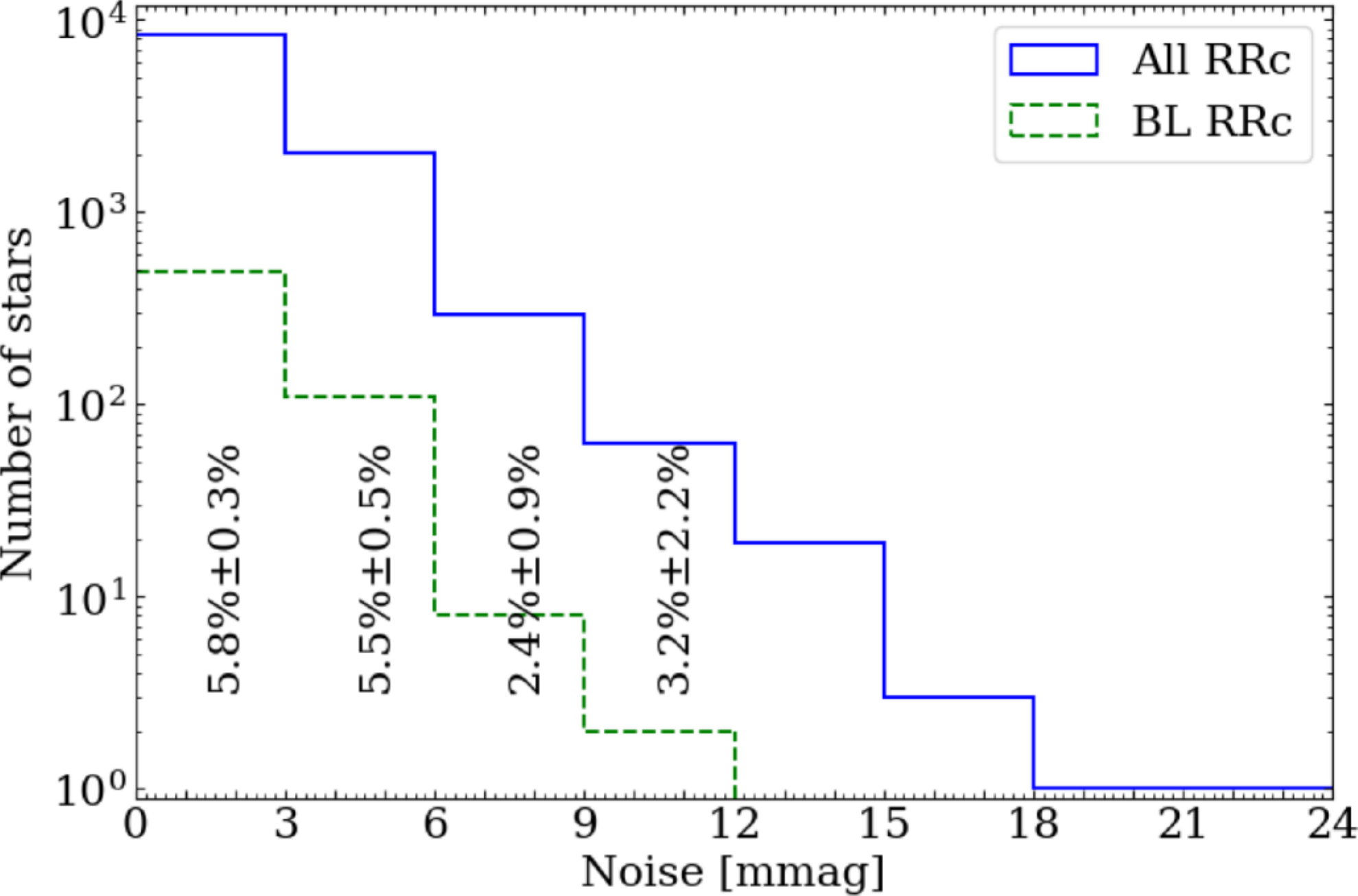}}
\caption{Distribution of noise level in power spectra of all RRc stars and all Blazhko stars. Incidence rate of Blazhko stars together with its statistical error is given for each bin. Statistical errors are calculated assuming that the population follows the Poisson distribution \citep[e.g.][]{alcockrrab}.}
\label{fig.noise}
\end{figure}

If we compare distributions of mean brightness for all RRc stars and for Blazhko RRc stars, we observe that Blazhko stars are slightly brighter. This effect is not very striking and is most likely caused by observational bias. For brighter stars photometric errors are lower. As a result the noise level in the power spectrum is lower and it is possible to detect modulation with lower amplitudes than for stars with higher noise level. The dependence of the incidence rate of the Blazhko effect on the noise level is studied in Fig.~\ref{fig.noise}. In order to estimate the noise level in the power spectrum of each star, we prewhitened the data with frequency of the first overtone and 7 harmonics. Then, the noise level is estimated as average signal between 6th and 7th harmonic as in this frequency range we do not expect any additional signals. Daily aliases of signals arising due to possible trends and period changes are also negligible. In Fig.~\ref{fig.noise}, distributions for all RRc stars and for Blazhko stars are plotted with solid and dashed lines, respectively. In the plot we provided incidence rates of Blazhko stars for each bin, together with statistical errors. As expected, the lower the noise level in the data, the higher the incidence rate of the Blazhko effect. We did not detect the Blazhko effect in stars with noise level higher than 10\,mmag.

We also compare light curve shapes of Blazhko and non-Blazhko stars. The best way to do it quantitatively, is to use the Fourier coefficients \citep{fcoeff}, amplitude ratios,
$$R_{k1}=A_k/A_1\,,$$
and phase differences,
$$\phi_{k1}=\phi_k-k\phi_1\,.$$

We investigated Fourier coefficients $R_{21}$, $\phi_{21}$, $R_{31}$ and $\phi_{31}$ as functions of the first-overtone period. Fourier coefficients for Blazhko stars were determined in this analysis. For each Blazhko star, the amplitudes and phases, that enter the above formulae, were taken from the Fourier fit, which included pulsation frequencies and modulation sidepeaks. Blazhko stars do not group around any specific values of the Fourier coefficients. They do not form any distinct progression as compared to non-modulated RRc stars. BLm stars are spread over similar period range as non-Blazhko RRc stars and their Fourier coefficients cover the same range as coefficients for non-Blazhko RRc stars. BL1 stars tend to have shorter pulsation periods but Fourier coefficients cover similar ranges as other RRc stars.

\cite{rrabbl} noticed a grouping of Blazhko RRab stars in Fourier parameter plots, especially for $R_{31}$ vs. $\phi_{31}$ and $\phi_{21}$ vs. $R_{31}$ plots. Based on these plots, they proposed a criterion to select Blazhko candidate stars. We also checked analogous plots for RRc stars, but there is no significant difference between Blazhko and non-Blazhko RRc stars' distribution. We note that light curve shapes for RRc stars are more sinusoidal than of RRab stars, and hence the Fourier parameter plots are, in general, more blurred and less structured.

\section{Period doubling in Blazhko RRc stars}\label{sec:pd}

One of the most intriguing phenomenons detected in Blazhko RRab stars is period doubling. In the light curve, it manifests as alternating minima and maxima during some phases of the Blazhko modulation. In the power spectrum, it manifests as signals located close to half-integers frequencies, i.e. signals at $\frac{2n+1}{2}\fo$. Period doubling was discovered for the first time in Blazhko RRab star RR~Lyr based on {\it Kepler} space photometry \citep{rrlyrpd}. Later studies revealed that many modulated RRab stars also show period doubling phenomenon \citep{benko2, szabo_doubling, szabo_doubling2}. Interestingly, no non-modulated RRab star was found to exhibit period doubling \citep{nonblrrab}. It suggests that there might be a connection between the Blazhko effect and period doubling \citep{szabo_doubling}.

\begin{figure}
\centering
\resizebox{\hsize}{!}{\includegraphics[bb= 0 0 590 383]{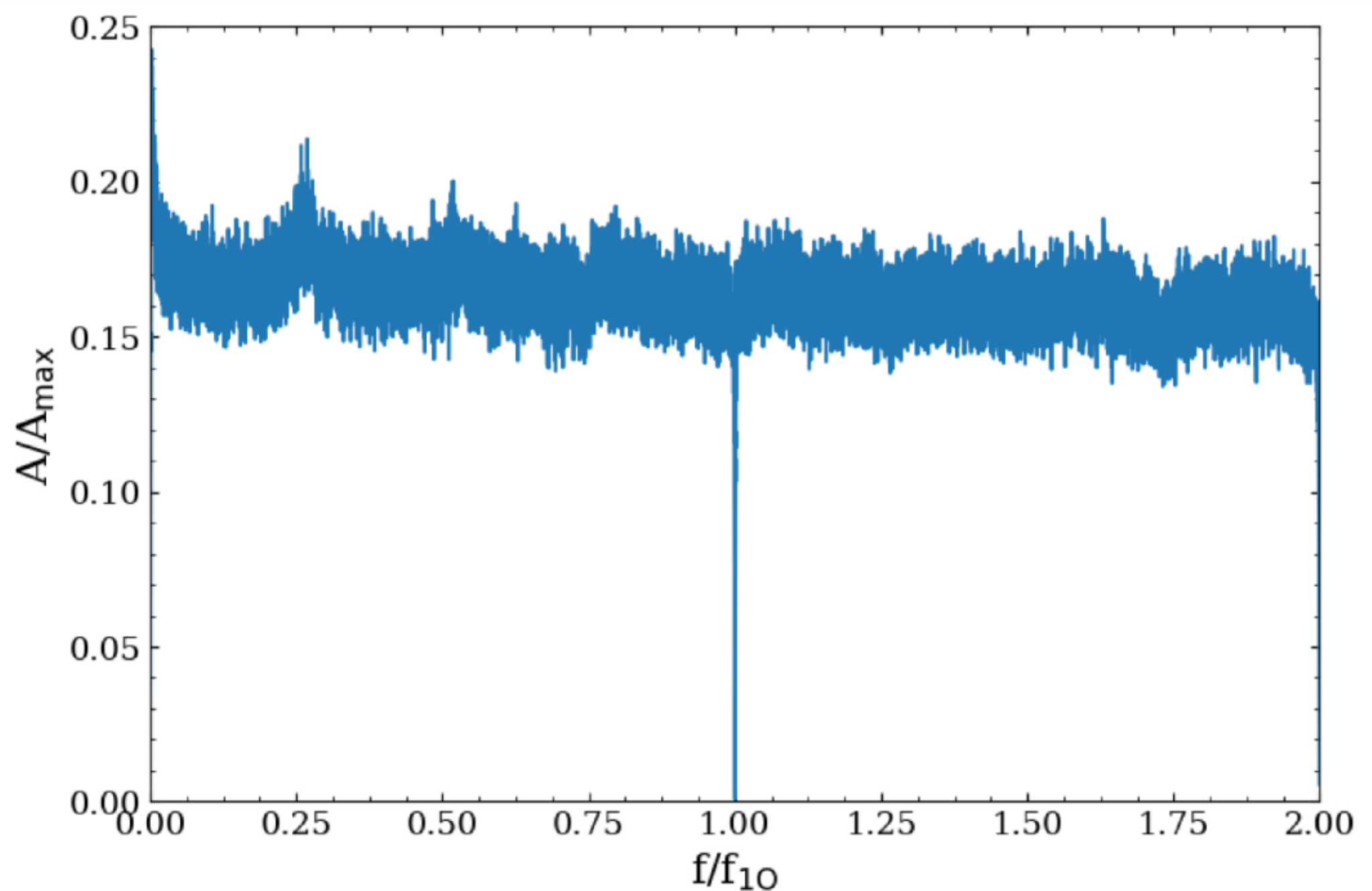}}
\caption{Average of the scaled power spectra for individual Blazhko stars resulting from the time-dependent prewhitening with first overtone and its harmonics. Spectra were scaled, so 1.0 corresponds to first overtone frequency in each star.}
\label{fig.av_spectrum}
\end{figure}

To investigate this problem in Blazhko RRc stars, we conducted a search for signals at half-integer frequencies among stars from our sample. During manual analysis of Blazhko stars we did not find significant signals located close to these frequencies. In order to check whether signal at these frequencies might be present in Blazhko RRc stars, we followed the procedure described in \cite{alcockrrab}, and averaged power spectra of Blazhko stars prewhitened with the dominant frequency and its harmonics. The result of this procedure, an averaged spectrum, is presented in Fig.~\ref{fig.av_spectrum}. All spectra were scaled before averaging, so that the first overtone frequency is located at $1$. In order to take into account different quality of spectra, we used the weighted arithmetic mean, where weights correspond to the inverse of the average noise level. Instead of standard prewhitening, we used a time-dependent version of the procedure, as described by \cite{068kepler}. This is because in many RRc stars we have to deal with period change of the first overtone, which results in a remnant power in the prewhitened spectrum at $\fo$ (Sect.~\ref{sec.analysis}). It increases the noise level and thus can hamper the detection of possible signals at half-integer frequencies. Time-dependent prewhitening was applied on a season-to-season basis. It resulted in a sharp drop of signal at $f/\fo=1$. While the method also removes slow trends from the data, possible variations within individual seasons are not removed and give rise to a signal (power excess) at low frequencies ($f/f_{\rm 1O}\approx 0$) and another power excess at $f/f_{\rm 1O}\approx 0.27$ (daily alias).

 If signals at half-integer frequencies are present in Blazhko RRc stars, we would detect them at frequencies $f/\fo=0.5$ or $1.5$ in Fig.~\ref{fig.av_spectrum}, which is not the case. We note, however, that detection of period doubling from ground-based photometry is challenging due to periods of RR Lyrae which make it hardly possible to cover two consecutive cycles during a single night. Moreover, In the case of RRab stars, sub-harmonics of the dominant modes often show offset from the exact subharmonic position, $\frac{2n+1}{2}\fo$, which was already reported by \cite{benko2} and \cite{szabo_doubling}. If it is the case also for RRc stars, it would hamper the detection of signals at half-integer frequencies in the averaged spectrum. This method is expected to amplify the low amplitude signals, hidden in the noise in individual frequency spectra, but it is possible only for signals located at the exactly same positions. Consequently, our results do not exclude the possibility that half-integer frequencies are present in modulated RRc stars.

\section{Results for individual stars}\label{sec.individual}

In this section we present analysis of the most interesting stars in our sample.

\subsection{OGLE-BLG-RRLYR-24030}\label{ssec:24030}

RRLYR-24030 is particularly interesting, because in this star we detected the shortest Blazhko period known so far, which is $2.23360(3)$\,d. In this star we detected first overtone frequency and its three harmonics. It is a BLm star with sidepeaks present only on the higher frequency side of the first overtone and its harmonics: $\fo+\fmod$, $2\fo+\fmod$ and $2\fo+2\fmod$ (incomplete quintuplet at the harmonic). Additionally, we detected modulation frequency in the low frequency range, $\fmod$. This star has also the highest pulsation amplitude in the considered sample, which exceeds 0.5\,mag. This star is certainly unusual in our sample, but detected signals do not fit to any other known group of double-periodic RR Lyrae stars and are within adopted criteria for Blazhko stars.

\subsection{OGLE-BLG-RRLYR-12500}

RRLYR-12500 is a BLm star with a very rich pattern of modulation sidepeaks placed symmetrically around $\fo$, and a short modulation period of $4.8892(1)$\,d. It is slightly shorter than the shortest Blazhko period for RRc star known so far, which is above $5$\,d \citep{skarka}. We detected several stars with Blazhko period below $5$\,d, but besides RRLYR-12500 other are BL1 stars or BLm stars with sidepeaks present only on one side of the first overtone frequency (like in RRLYR-24030 discussed above).

\begin{figure}
\centering
\resizebox{\hsize}{!}{\includegraphics{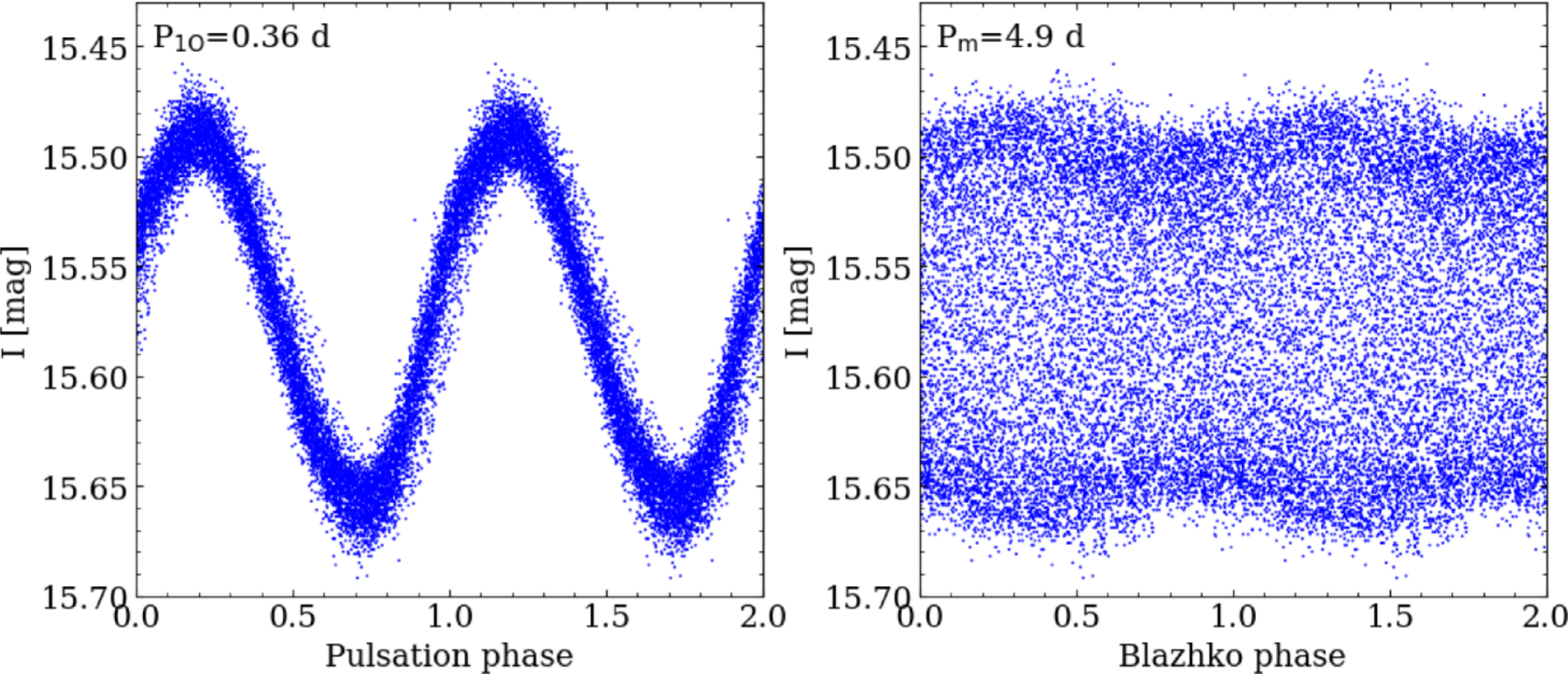}}
\caption{Light curve of RRLYR-12500 phased with the first overtone period (left panel) and with the modulation period (right panel).}
\label{fig.lcf_12500}
\end{figure}

In Fig.~\ref{fig.lcf_12500} we present data for RRLYR-12500 phased with the first overtone frequency and with the modulation frequency (left and right panels, respectively). Blazhko effect in RRLYR-12500 is not pronounced -- modulation is clearly visible in the right panel of Fig.~\ref{fig.lcf_12500}, but is of low amplitude. Relative modulation amplitude is only 8\,per cent and data phased with pulsation period show relatively small scatter.

\begin{figure}
\centering
\resizebox{\hsize}{!}{\includegraphics{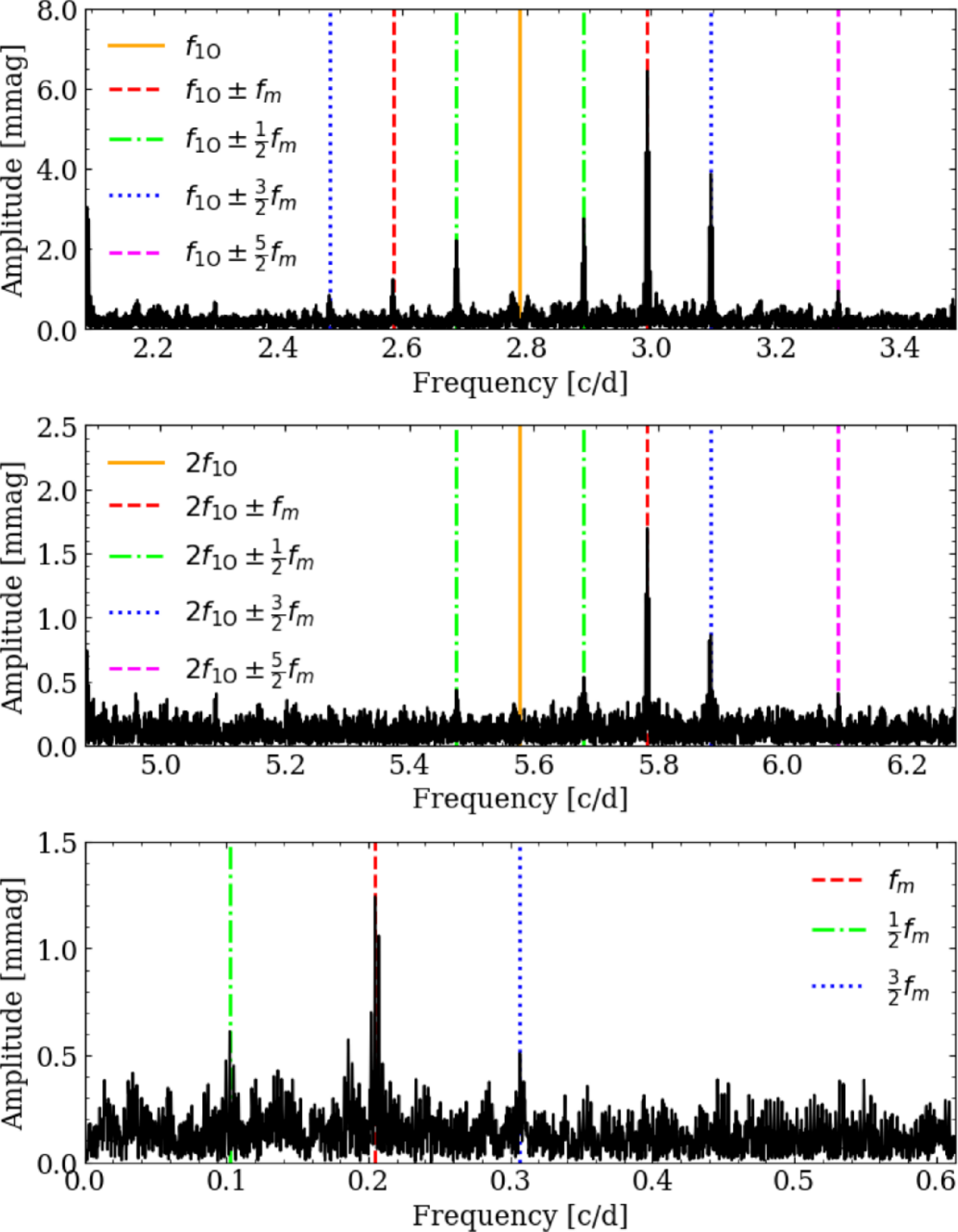}}
\caption{Power spectra of RRLYR-12500 after prewhitening with the first overtone frequency and its three harmonics. Top panel shows vicinity of the first overtone frequency. Middle panel shows vicinity of the first harmonic. Positions of the first overtone and its harmonic are marked with orange solid line. Modulation sidepeaks are plotted with red dashed line. Subharmonics of the modulation frequency are plotted with different line styles and colours indicated in the key of the figure. Bottom panel shows the low frequency range of the spectrum. Modulation frequency is plotted with red dashed line. Subharmonic of modulation frequency are plotted with the same styles and colours as in the panels above.}
\label{fig.fs_12500}
\end{figure}

Despite low modulation amplitude, the frequency spectrum of this star, illustrated in Fig.~\ref{fig.fs_12500}, is particularly rich and interesting, with an uncommon feature. Top panel of Fig.~\ref{fig.fs_12500} shows the vicinity of the prewhitened $\fo$, middle panel shows the vicinity of the prewhitened first harmonic (solid orange lines), and bottom panel shows the low frequency range. Detected modulation sidepeaks are marked with vertical lines of different colours and line-styles. Besides the triplet components, $\fo\pm\fmod$, and $2\fo+\fmod$ (red dashed lines), we detected a few additional signals of lower amplitude, which can be explained as subharmonics of the Blazhko components. These are $\fo\pm \frac{1}{2}\fmod$, and $2\fo\pm\frac{1}{2}\fmod$ (green dot-dashed lines), $\fo\pm\frac{3}{2}\fmod$, and $2\fo+\frac{3}{2}\fmod$ (blue dotted lines) and even $\fo+\frac{5}{2}\fmod$ and $2\fo+\frac{5}{2}\fmod$ (magenta dashed lines). This is very uncommon situation for a Blazhko star, so far noticed in a few objects only. Subharmonics of modulation frequency were detected for the first time in the multiperiodic Blazhko star CZ~Lac by \cite{sodor}. \cite{benko2} also found similar structures in the power spectra of four stars observed by {\it Kepler}. The same behaviour was found in two anomalous RRd stars by \cite{rrdbl}. As discussed in these papers, the subharmonic frequency, $\frac{1}{2}\fmod$, could be a real modulation frequency, $\fmod'=\frac{1}{2}\fmod$. However, in this scenario the sidepeaks at $\fo \pm 2\fmod'$ would have higher amplitude than the triplet components, $\fo \pm \fmod'$, and a modulation sidepeak of a very high order, at $\fo+5\fmod'$ would be detected, which seems unlikely. Moreover, in the low frequency range we observe a similar situation (the bottom panel of Fig.~\ref{fig.fs_12500}). Signal at $\fmod$ has the highest amplitude, but we also detected two signals with significantly lower amplitudes, at $\frac{1}{2}\fmod$, and at $\frac{3}{2}\fmod$.

In analogy with the period doubling effect known in type II Cepheids or in the Blazhko RRab stars, in which the consecutive pulsation cycles differ, we may deal with the period-doubled modulation cycle here. The modulation pattern would repeat after two modulation periods then. Besides RRLYR-12500, we found signals corresponding to $\fo\pm \frac{1}{2}\fmod$ in 5 additional stars. In Tab.~A1 they are marked with `e' in the remarks column.
 
\subsection{OGLE-BLG-RRLYR-11167}

\begin{figure}
\centering
\resizebox{\hsize}{!}{\includegraphics{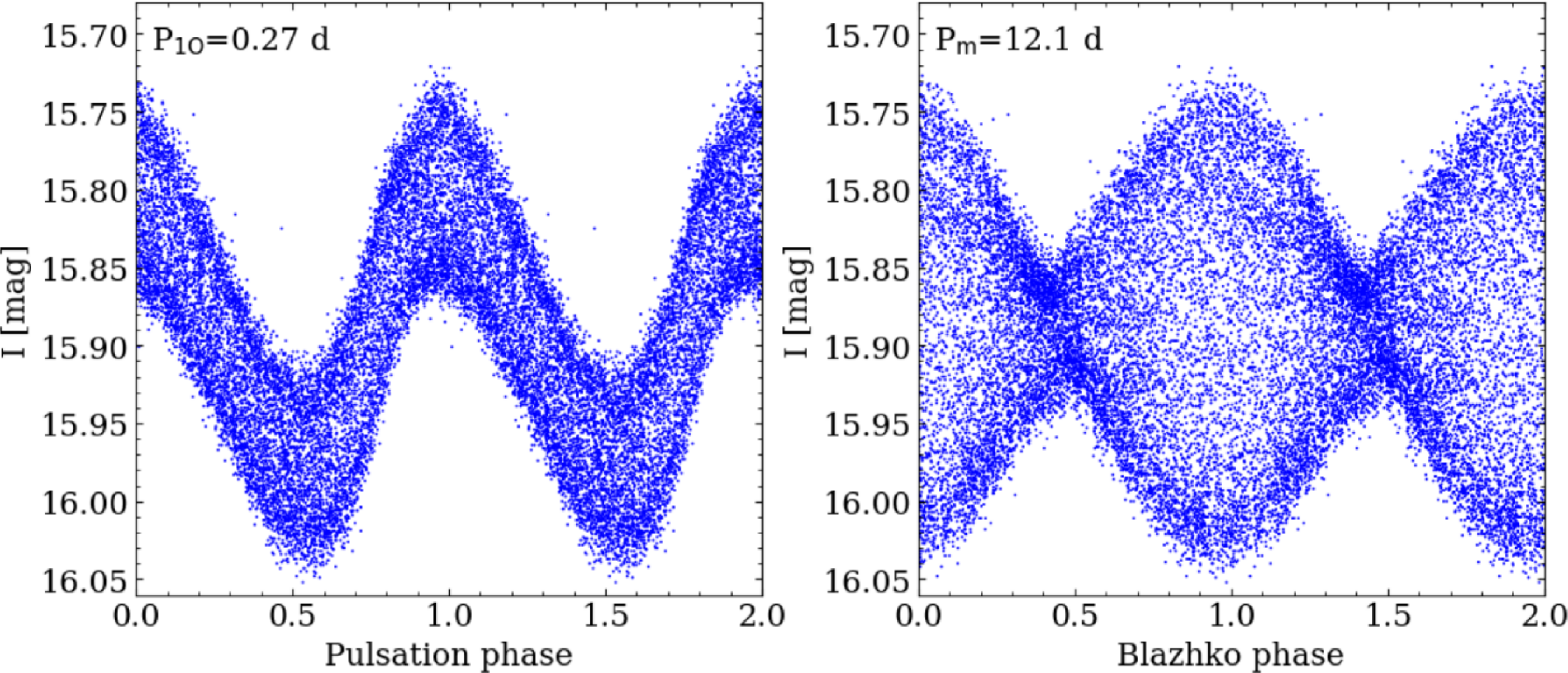}}
\caption{Light curve of RRLYR-11167 phased with the first overtone period (left panel) and with modulation period (right panel).}
\label{fig.lcf_11167}
\end{figure}

In RRLYR-11167 we detected Blazhko effect with period $P_{\rm m}=12.1324(1)$\,d. Data phased with the first overtone period are presented in the left panel of Fig.~\ref{fig.lcf_11167}, and show significant scatter across pulsation phases, which indicates Blazhko effect with high relative modulation amplitude. Indeed, $A_-/A_{\rm 1O}$, is 62\,per cent. This is very well visible when we fold data with the Blazhko period, which is shown in the right panel of Fig.~\ref{fig.lcf_11167}. 

\begin{figure*}
\centering
\resizebox{\hsize}{!}{\includegraphics{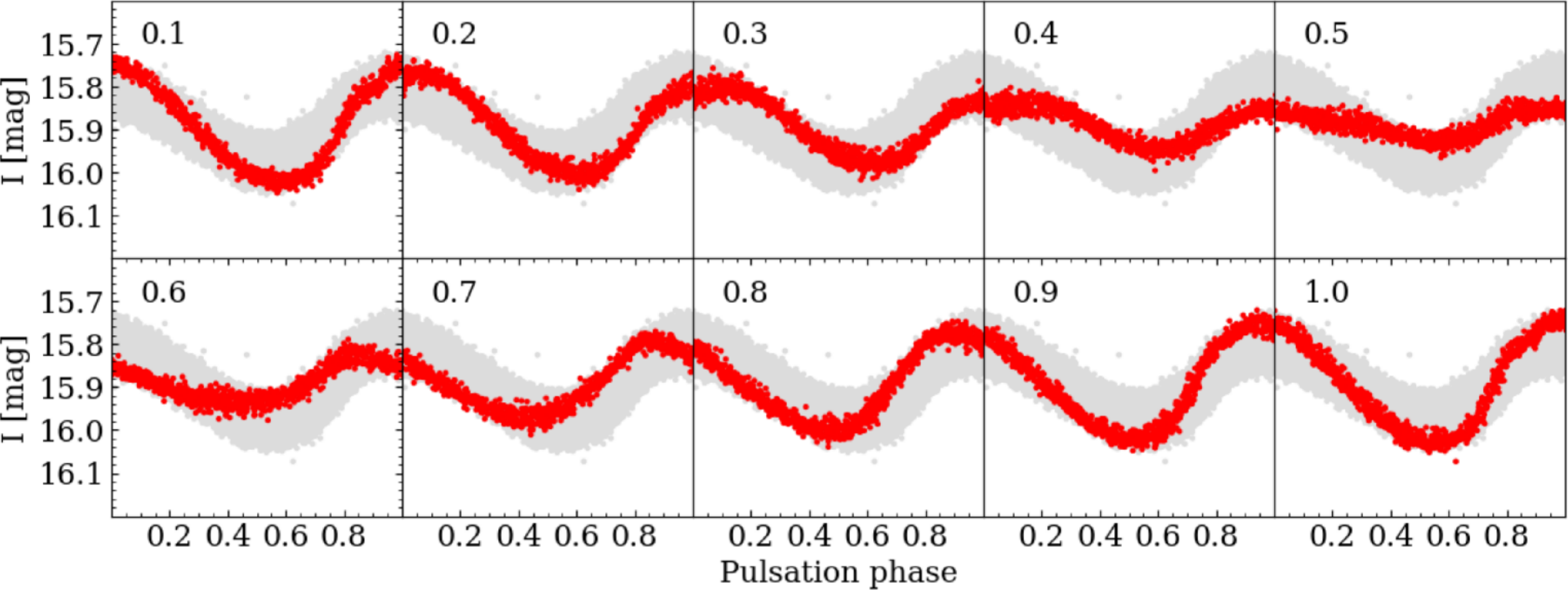}}
\caption{Light curves of RRLYR-11167 phased with the pulsation period for different Blazhko phases. Blazhko phase is indicated in the top left corner of each panel. Grey points correspond to the whole dataset. Red points correspond to subsets from each Blazhko phase.}
\label{fig.blph_11167}
\end{figure*}

For RRLYR-11167, short modulation period and numerous data (more than $10\,000$ observations) result in an excellent phase coverage of the Blazhko cycle and allows detailed  study of the light curve changes. Light curves for different Blazhko phases are shown in Fig.~\ref{fig.blph_11167}. Top left corner of each panel indicates Blazhko phase. Grey points correspond to whole data set phased with pulsation period. Data subsets, corresponding to different Blazhko phases, folded with pulsation period, are plotted with red points. Highest amplitude of pulsation is observed for Blazhko phases around 0, and the smallest pulsation amplitude is observed for Blazhko phases close to 0.5.
 
\begin{figure}
\centering
\resizebox{\hsize}{!}{\includegraphics{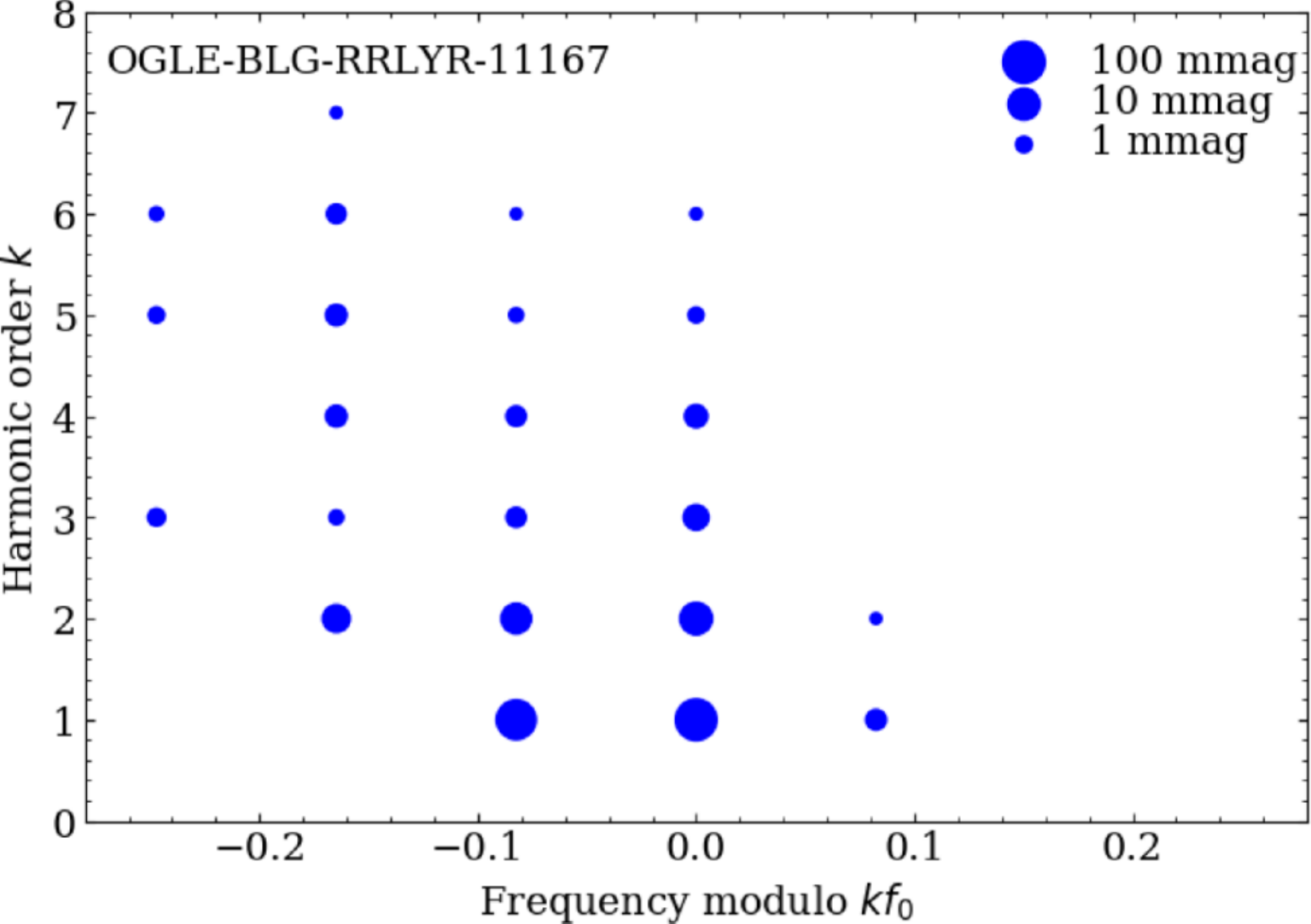}}
\caption{Echelle diagram for RRLYR-11167. Size of the points corresponds to the amplitude of the signal, as indicated in the key.}
\label{fig.echelle_11167}
\end{figure}

In the power spectrum of RRLYR-11167 we detected 5 harmonics of the first overtone frequency. Triplet components are present at the frequency of the first overtone and its first harmonic. For higher harmonics we detected components of quintuplets and septuplets, but only on the low-frequency side. Detected frequencies are presented in Fig.~\ref{fig.echelle_11167} in the form of an echelle diagram \citep{guggenberger2012}. Size of a point in Fig.~\ref{fig.echelle_11167} indicates the amplitude of a corresponding signal.

After prewhitening with frequency of the first overtone, its harmonics and Blazhko multiplets, we detected two additional close signals at $f_{\rm x}/\fo=0.68271$ (S/N=13.22) and at  $f_{\rm y}/\fo=0.68106$ (S/N=9.02). Thus the star may be classified as a member of $RR_{0.68}$ group, although we note that the period ratios are a bit smaller than typically observed in $RR_{0.68}$ stars, which tightly cluster around period ratio $0.686$ \citep{068}. In addition, in all known stars observed in this group we detect a single periodicity. The peak at $f_{\rm y}$ may be interpreted as arising due to modulation of a periodicity at $f_{\rm x}$. The modulation period would be $162$\,d, then, which is different than the Blazhko period.

\subsection{Double-periodic Blazhko star OGLE-BLG-RRLYR-03825}

\begin{figure}
\centering
\resizebox{\hsize}{!}{\includegraphics{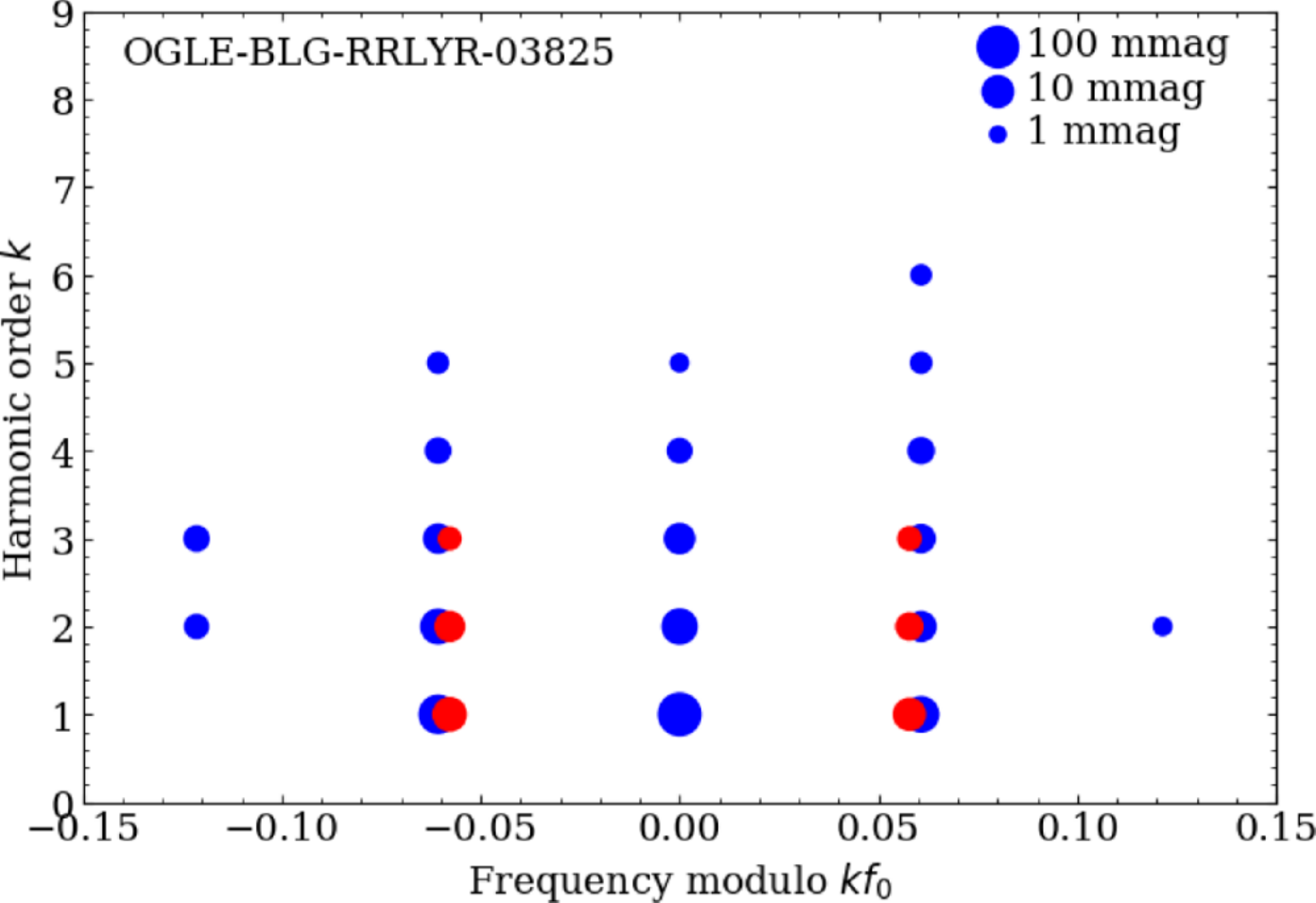}}
\caption{Echelle diagram for RRLYR-03825. Blue points correspond to the primary modulation, red points -- to the secondary modulation. Size of the points indicates amplitude of the corresponding signal.}
\label{fig.echelle_03825}
\end{figure}

In RRLYR-03825 we detected two Blazhko periods $P_{\rm m_1}=16.4704(4)$\,d and $P_{\rm m_2}=17.304(1)$\,d. All detected signals are presented in Fig.~\ref{fig.echelle_03825} in the form of an echelle diagram. 

We detected harmonics of the first overtone frequency up to $5\fo$ and at all these harmonics we observe triplets corresponding to the Blazhko period $P_{\rm m_1}$. At $2\fo$ we detected quintuplet and at $3\fo$ we detected one component of quintuplet. Secondary modulation, $P_{\rm m_2}$, gives rise to three triplets in the power spectrum, which are located at the frequency of the first overtone and its two harmonics. 

 Relative modulation amplitudes for these two modulation periods, $P_{\rm m_1}$ and $P_{\rm m_2}$, are 31\,per cent and 11\,per cent, respectively. In the low-frequency range we detected signals corresponding to dominant modulation frequency, $f_{\rm m_1}$, and its harmonic, $2f_{\rm m_1}$.

\subsection{Three triple-periodic Blazhko stars}\label{ssec:threemod}

\begin{figure}
\centering
\resizebox{\hsize}{!}{\includegraphics{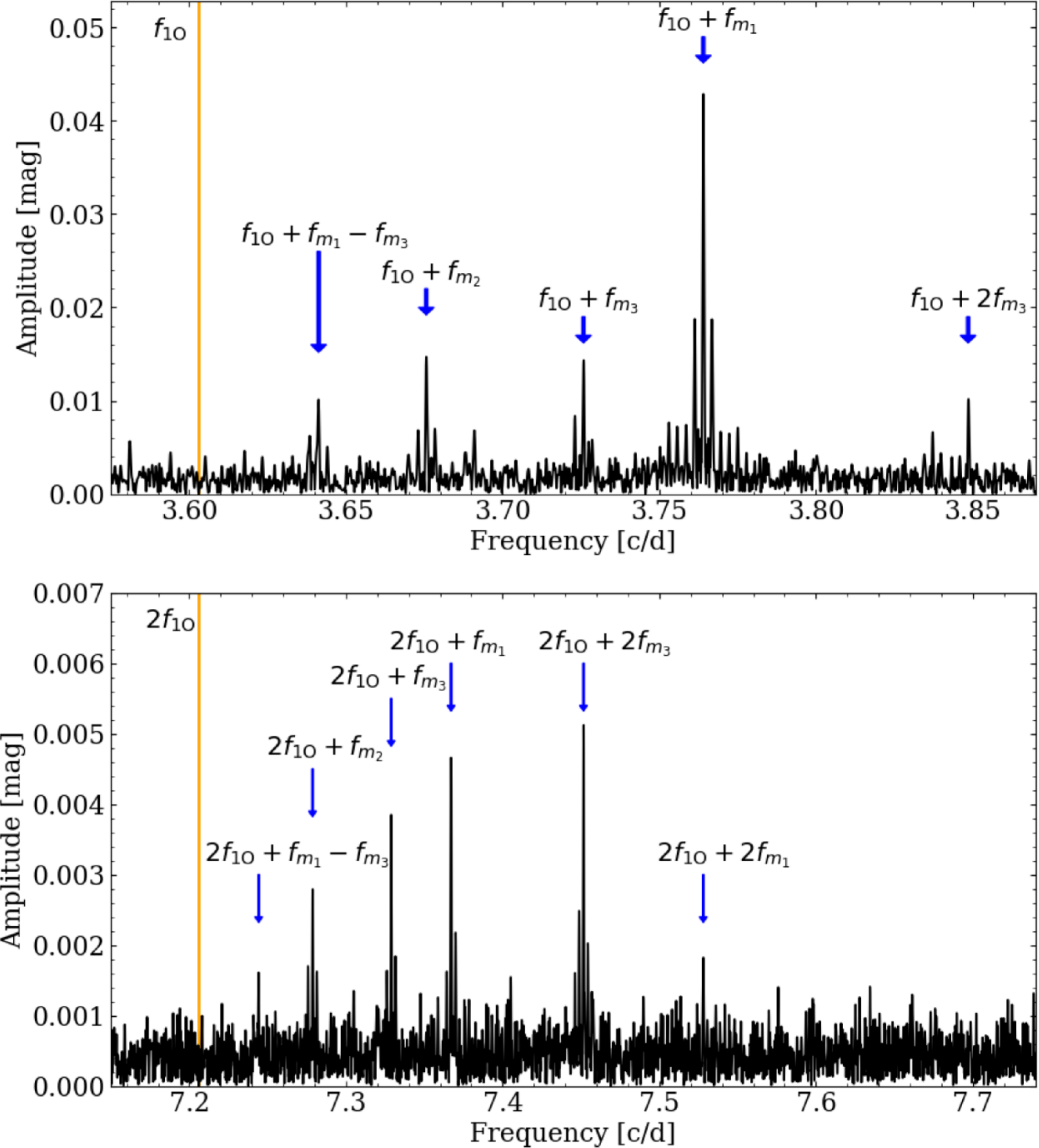}}
\caption{Power spectra for RRLYR-33726 after prewhitening with the first overtone and its harmonic (orange solid line). Top panel shows vicinity of the first overtone, bottom panel -- of its harmonic. Arrows indicate positions of Blazhko sidepeaks. Labels show their identifications.}
\label{fig.33726_fs}
\end{figure}

In OGLE-BLG-RRLYR-33726 we detected three Blazhko periods: $P_{\rm m_1}=6.21156(8)$\,d, $P_{\rm m_2}=13.777(1)$\,d and $P_{\rm m_3}=8.146(2)$\,d. In the power spectrum, we detected frequency of the first overtone and its first harmonic. Besides peaks corresponding to pulsation in the first overtone mode, we detected sidepeaks corresponding to three separate Blazhko periods and their linear combinations. Interestingly, the only sidepeaks that we detected are located on the higher-frequency side of $k\fo$. In Fig.~\ref{fig.33726_fs} we plot the power spectrum after prewhitening with the frequency of the first overtone and its harmonic. Top panel shows vicinity of the first overtone frequency. Its position is marked with orange solid line. Bottom panel is plotted in the same way, but for the vicinity of the first harmonic. Arrows indicate modulation sidepeaks with identifications included in the plot. Additionally, we detected modulation frequencies in the low frequency range corresponding to $P_{\rm m_1}$ and $P_{\rm m_2}$.

 The second star in which we detected three Blazhko periods is OGLE-BLG-RRLYR-06207. Three detected periods are: $P_{\rm mod_1}=66.18(2)$\,d, $P_{\rm m_2}=7.4955(3)$\,d and $P_{\rm m_3}=8.875(1)$\,d. For this star we have more than $12\,000$ observations, so the noise level in the power spectrum is low. In power spectrum we detected up to seven harmonics of the first overtone. For the primary Blazhko period we detected only dublets at the frequency of the first overtone and its first harmonic. For other Blazhko periods we detected triplets. In contrast to RRLYR-33726 we did not detect any higher-order sidepeaks or frequencies of modulation in the low frequency range or combination frequencies between sidepeaks.
 
The third star with three Blazhko periods is OGLE-BLG-RRLYR-06625. Three detected periods are: $P_{\rm m_1}=5.0407(2)$\,d, $P_{\rm m_2}=39.23(2)$\,d and $P_{\rm m_3}=29.24(2)$\,d. All three modulations give rise to triplets in the frequency spectrum at the frequency of the first overtone. At the harmonic, only triplets corresponding to $P_{\rm m_1}$ and $P_{\rm m_2}$ are detected.

\subsection{Star with four Blazhko periods}\label{ssec:fourmod}
 
RRLYR-32935 has a complicated power spectrum with many signals. At first overtone frequency we detect four distinctive signals corresponding to one triplet and three dublets. Periods of modulations are $P_{\rm m_1}=4.28326(4)$\,d, $P_{\rm m_2}=460.3(4)$\,d, $P_{\rm m_3}=22.120(1)$\,d and $P_{\rm m_4}=4.25260(6)$\,d. The only triplet is due to modulation with the longest period, $P_{\rm m_2}$. $P_{\rm m_1}$ and $P_{\rm m_4}$ are very close, but resolved, hence, we included them in the solution as two independent frequencies. At $2\fo$ we detected dublets corresponding to all modulation periods. Besides these structures we detected many signals which can be explained as linear frequency combinations between modulation frequencies, e.g. $2\fo+f_{\rm m_1}+f_{\rm m_2}$, $2\fo+f_{\rm m_1}-f_{\rm m_3}$ etc.
 
The four-period Blazhko modulation in RRLYR-32935 is somewhat suspicious. The power spectrum has many signals, but we observe equidistant triplet for one modulation period only. Otherwise, we observe doublets and many combination signals between Blazhko sidepeaks.

Another possible explanation for RRLYR-32935 is that it is in fact a multiperiodic $\delta$~Scuti star with three independent close frequencies, linear combinations between them and modulation of the dominant frequency. The frequency spectrum is indeed similar to that computed for Galactic bulge multi-periodic $\delta$~Scuti stars (Netzel et al., in prep.), but the pulsation periods would be rather long then. Fourier decomposition parameters for RRLYR-32935 are lower than typical for RRc stars, but still within observed values for these stars. In Tab.~\ref{tab.4per} we list all signals detected in the power spectrum of RRLYR-32935 with their frequencies, amplitudes and phases, and two possible interpretations: Blazhko RRc (fourth column) and multiperiodic $\delta$~Scuti-type star (fifth column). While the frequencies listed in the last column -- low order linear combinations -- are commonly detected in multiperiodic stars, linear combinations between modulation frequencies, as numerous as in RRLYR-32935, are not detected in any other Blazhko RRc we have analysed. 

\begin{table*}
\centering
\caption{Frequencies detected in power spectrum of RRLYR-32935. Consecutive columns provide frequency, its amplitude, phase and two possible explanations of the signal.}
\label{tab.4per}
\begin{tabular}{@{}||lllll@{}}
Frequency [c/d] & $A$ [mag] & $\phi$ [rad] & Blazhko RRc & $\delta$~Scuti \\
\hline
\hline
3.91519195 & 0.0636 & 3.66 & $\fo$	&	$f_0$ \\
7.83038390 & 0.0057 & 3.53 & $2\fo$	&	$2f_0$ \\
4.14865898 & 0.0225 & 2.96 & $\fo+f_{\rm m_1}$	&	$f_{\rm x}$ \\
3.91736426 & 0.0187 & 1.81 & $\fo+f_{\rm m_2}$	&	$f_0+\fmod$ \\
3.86998343 & 0.0164 & 0.22 & $\fo-f_{\rm m_3}$	&	$f_{\rm y}$ \\
3.91301964 & 0.0145 & 2.48 & $\fo-f_{\rm m_2}$	&	$f_0-\fmod$ \\
4.15034216 & 0.0135 & 4.51 & $\fo+f_{\rm m_4}$	&	$f_{\rm z}$ \\
8.06602324 & 0.0044 & 6.28 & $2\fo+f_{\rm m_1}+f_{\rm m_2}$	&	$f_0+f_{\rm x}+\fmod$ \\
8.01864241 & 0.0033 & 4.66 & $2\fo+f_{\rm m_1}-f_{\rm m_3}$	&	$f_{\rm x}+f_{\rm y}$ \\
7.83255621 & 0.0023 & 2.07 & $2\fo+f_{\rm m_2}$	&	$2f_0+\fmod$ \\
7.78300307 & 0.0027 & 5.48 & $2\fo-f_{\rm m_2}-f_{\rm m_3}$	&	$f_0+f_{\rm y}-\fmod$ \\
8.06385093 & 0.0023 & 2.86 & $2\fo+f_{\rm m_1}$	&	$f_0+f_{\rm x}$ \\
0.23346703 & 0.0025 & 0.85 & $f_{\rm m_1}$	&	$f_{\rm x}-f_0$ \\
0.04520852 & 0.0016 & 5.19 & $f_{\rm m_3}$	&	$f_0-f_{\rm y}$ \\
3.86781112 & 0.0017 & 0.01 & $\fo-f_{\rm m_2}-f_{\rm m_3}$	&	$f_{\rm y}-\fmod$ \\
7.82821159 & 0.0014 & 2.52 & $2\fo-f_{\rm m_2}$	&	$2f_0-\fmod$ \\
8.29900115 & 0.0015 & 1.58 & $2\fo+f_{\rm m_1}+f_{\rm m_4}$	&	$f_{\rm x}+f_{\rm z}$ \\
8.06553411 & 0.0012 & 4.48 & $2\fo+f_{\rm m_4}$	&	$f_0+f_{\rm z}$ \\
7.78517538 & 0.0011 & 0.06 & $2\fo-f_{\rm m_3}$	&	$f_0+f_{\rm y}$ \\
\end{tabular}
\end{table*}

\label{sec.individual}

\section{Discussion}\label{sec.discussion}

\subsection{Incidence rate of the Blazhko stars}\label{ssec:incrate}

In Tab.~\ref{tab.rates} we present the incidence rates of the Blazhko effect in RRab and RRc stars in various stellar systems. RRc stars from the Galactic bulge were previously analysed based on OGLE-I and OGLE-II data by \cite{moskalikporetti} and \cite{mizerski}, respectively. \cite{moskalikporetti} analyzed 65 RRc stars and found 3 Blazhko stars (two of BL1 type and one BLm star). The inferred incidence rate based on their study is $4.6\%\pm2.6$\,per cent. \cite{mizerski} analyzed a sample of 771 RRc stars and found 52 Blazhko stars (22 BL1 and 30 BLm stars), which corresponds to incidence rate of $6.6\%\pm0.9$\,per cent. Despite the fact, that we used data with a significantly larger number of observations and of better quality, we did not find that the incidence rate of the Blazhko phenomenon is larger. We note however that in the study of \cite{mizerski} the criterion, which was used to classify a star as Blazhko, was not clearly spelled out (the list of modulated stars was not published as well), while the study by \cite{moskalikporetti} is based on a very small sample. Therefore, it is difficult to objectively compare the incidence rates.

The highest incidence rates of the Blazhko effect in RRc stars were reported in the globular clusters, M3 and NGC6362 (Tab.~\ref{tab.rates}). We note that these studies are based on the analysis of the top-quality data gathered during dedicated monitoring of these clusters.


The OGLE-IV data were used  by \cite{rrabbl} for the analysis of the Blazhko effect in RRab stars from the Galactic bulge. The inferred incidence rate of Blazhko stars is $40.3$\,per cent. Again incidence rate for RRab stars is significantly higher than for RRc stars, as was already noticed by several studies and is summarized in Tab.~\ref{tab.rates}. We note that \cite{rrabbl} did not analyse the full sample of RRab stars, but used additional criterions to select a high-quality sample. Two of the criterions where number of data points and mean brightness. Only stars brighter than $18$\,mag, for which more than $420$ data points were available, were analysed. Obviously, for brighter stars with more data points, one may expect lower noise level and consequently higher chances to detect signals of low amplitude (see also Fig.~\ref{fig.noise} and discussion in Sect.~\ref{sec:blvsnbl}). If we apply these two criterions to our RRc sample, we are left with $5\,860$ stars. Among them $446$ are Blazhko stars, which constitutes $7.6$\,per cent. As expected, this incidence rate is higher than that obtained for the whole sample of RRc stars ($5.6$\,per cent). We conclude that in the Galactic bulge the Blazhko effect is more than 5 times less frequent in RRc stars than in RRab stars.

\begin{table*}
\centering
\caption{Incidence rates of the Blazhko effect in RRab and RRc stars in various stellar systems.}
\label{tab.rates}
\begin{tabular}{@{}||lclcl@{}}
  
 system  & RRab  & Reference & RRc & Reference \\
   \hline
   \hline
   Galactic bulge & 23\% & \cite{moskalikporetti} & 5\% & \cite{moskalikporetti} \\
   & 20\% & \cite{mizerski} & 7\% & \cite{mizerski} \\
    & 40.3\% & \cite{rrabbl} & 5.6\% & this study \\
   LMC & 12\% & \cite{alcockrrab} & 4\% & \cite{alcockrrc}  \\
      &       &                   & 7.5\% & \cite{nagykovacs}  \\
   SMC & 10\% & \cite{smcbl} & 10\% & \cite{smcbl} \\
   M3 & 50\% & \cite{jurcsik3} & 12\% & \cite{jurcsik3}  \\
   Galactic field & 47\% & \cite{jurcsik2} & & \\
   NGC 6362 & 69\% & \cite{ngc} & 19\% & \cite{ngc} \\
  
\end{tabular}
\end{table*}

\subsection{RRc vs. RRab}

The incidence rate is not the only difference between Blazhko RRab and RRc stars. For BLm stars we have calculated the asymmetry parameter, $Q$. The distribution is asymmetrical with average value equal to $-0.1$. It means that sidepeaks located on low-frequency side are dominant, which is in contrast to what was found for modulated RRab stars. \cite{alcockrrab} calculated this parameter for a numerous sample of RRab stars based on MACHO LMC data. For majority of stars they found $Q$ in between 0.1 and 0.6 and the distribution peaks at 0.3 -- see their fig.~10. This is in an agreement in what \cite{mizerski} observed for modulated RR~Lyrae stars in the Galactic bulge. According to his analysis, RRab stars tend to have dominant sidepeaks on the high-frequency side, while in RRc stars he observed the opposite preference.

Otherwise, the overall picture of modulation in the frequency domain is very similar for RRab and RRc stars: we detected multiplet structures (quintuplet and septuplet components), and signature of mean brightness modulation, i.e. peaks at the modulation frequency. Some peculiarities reported for Blazhko RRab stars, like sub-harmonics of the modulation frequency, e.g. peaks at $\fo\pm\frac{1}{2}\fmod$, were also detected in Blazhko RRc stars. We have also detected multiperiodic Blazhko effect. \cite{benko2} noted that ratios of the modulation periods in RRab stars are, in some cases, close to the ratios of small integer numbers. This is not the case for modulated RRc stars.

The period doubling phenomenon in the Blazhko RRab star was detected for the first time in the {\it Kepler} photometry for the RR~Lyr. Later analysis of the space data showed that the period doubling is present in the majority of modulated RRab stars -- see Sect.~\ref{sec:pd} for more detailed discussion. In some cases this effect is weak, but sometimes the amplitude of alternations can be as high as $0.1$\,mag. The reason, why it was not detected in ground-based observations is the typical period of pulsations in RRab stars, which is around $0.5$\,d. In principle, it should be more easy to detect this effect for RRc stars in the ground-based data, thanks to their shorter pulsation periods, provided the effect is also present in modulated RRc stars. However, in our study we did not see any signs of period doubling in the power spectrum during the manual analysis of these stars. We also averaged spectra of all Blazhko stars (Sect.~\ref{sec:pd}), but could not find a signature of the effect as well. This however, might by due to the fact that the signals arising from period doubling might not be located exactly at the subharmonic frequencies. 

The power spectra of Blazhko RRab stars observed by {\it CoRoT} and {\it Kepler} are often rich in additional, low amplitude periodicities. Most of these can be interpreted as due to excitation of radial second overtone, or as arising due to period doubling effect -- for a summary see \cite{garden}. Interestingly, non-modulated RRab stars appear very regular, void of additional low-amplitude signals \citep{nonblrrab,benkoszabo}. In modulated RRc stars we also observe a wealth of additional signals (Sect.~\ref{ssec:as}), which may correspond to both radial (anRRd) and non-radial modes, e.g. $RR_{0.61}$ stars. Contrary to RRab stars, additional low-amplitude signals are even more common in non-modulated RRc stars. Both in $RR_{0.61}$ and in $RR_{0.68}$ the dominant variability is due to radial first overtone and majority of these stars are non-modulated.

\subsection{BL1 vs BLm}

In the preceding sections we have pointed some differences between BL1 and BLm stars. BL1 stars tend to have shorter modulation periods. Higher relative modulation amplitudes are more common among BL1 stars. The signal at modulation frequency is more common among BL1 stars (46 per cent) than among BLm stars (24 per cent).

In principle, BL1 stars could be double-mode radial-non-radial pulsators. In such scenario, the sidepeak detected in the vicinity of radial first overtone would correspond to non-radial mode. Other sidepeaks detected at higher order harmonics, or in the low frequency range, would be simple linear combination frequencies then. We note that such interpretation is much more difficult for stars with triplet and multiplet structures in the frequency spectra, as then higher order linear combination frequencies must be invoked, which is atypical for double-mode pulsators. But also for BL1 stars, the non-radial mode scenario faces difficulties. Low-degree non-radial modes are excited in RR~Lyrae stars as analysed, e.g. by \cite{vdk98} and \cite{dc99}. Their spectrum is dense, i.e. there are many unstable modes of a given degree with close frequencies, but their growth rates are at least order of magnitude lower than for radial modes. There is no obvious mode selection mechanism at work. If we assume that the linear growth rate is a good predictor of mode's amplitude, which is not necessarily the case -- see, e.g. \cite{smolec14}, then one would expect the excitation of non-radial modes preferentially on particular side of the radial mode frequency, which is clearly not the case. The very high relative modulation amplitudes we observe in some of BL1 stars would be hard to explain as well.

Consequently, modulation seems a much more reliable scenario for BL1 stars. It is also a common assumption in the studies of the Blazhko effect that doublets are due to modulation. In general, modulation manifests as equally spaced multiplets in the frequency spectra. The amplitudes of the modulation sidepeaks may be highly asymmetric however -- they depend, e.g. on the relative strength and phase relation between the amplitude and phase modulations \citep{bsp11}. The accuracy of the observations are often too low to see the low amplitude component in the case of the BLm stars; therefore, they are often mis-classified as BL1 stars. We note that quite often data of better quality lead to the discovery of the triplet or multiplet components `missing' in the data of lower quality, which supports the above view. This is also the case for many stars analysed in the present paper using the OGLE-III and OGLE-IV data. Revisiting the stars analysed by \cite{moskalikporetti}, who used the OGLE-I data, leads to the same conclusions: many of their BL1 type stars turn to be BLm with newer data. A close to 90 degree lag between the amplitude and phase modulations may lead to extremely non symmetric patterns, that would reveal as strongly non-symmetric multiplets even in the top-quality data. In the present study we have detected not only BL1, i.e. doublet patterns, but also strongly asymmetric multiplet patterns, i.e. incomplete quintuplets with all modulation sidepeaks present on one side of the radial mode frequency.

\subsection{Origin of the Blazhko effect}

Since the discovery of the Blazhko effect over a hundred years ago, its origin remains a mystery. Moreover, we observe that the high fraction of RR~Lyrae stars show the phenomenon. Several models have been proposed for the Blazhko effect, but none is satisfactory. Majority of these models are focused on RRab stars, since the incidence rate is higher for these stars and they are more extensively studied from ground and from space. 

\cite{kovacs_cokon} (see his tab.~2) summarized the most significant observational properties of the Blazhko RRab stars that should be explained by the model. Some of these properties are also observed in Blazhko RRc stars, e.g. cases of high relative amplitude ratio, cases of strong asymmetry of amplitudes of the sidepeaks, range of modulation periods (in case of RRc stars in this study the range is $2-3000$\,d), stars with multiperiodic modulations. We also detected Blazhko RRc stars with additional modes, including non-radial ones.

Till recently, the most elaborated models were oblique magnetic rotator \citep{shibahashi} and resonant nonradial rotator/pulsator model \citep{nowakowski_dziembowski,dziembowskimizerski}. Both have the same drawbacks. They predict clock-work modulation in which modulation period is equal to the rotation period of a star. Hence, they fail to explain irregular modulation cycles and the very wide range of modulation periods, which in case of RRc stars vary from 2 to more than 2000\,d. According to these models, in the frequency domain, the modulation should manifest as symmetric triplets (resonant nonradial rotator) or symmetric quintuplets (oblique magnetic rotator). In both RRab and RRc stars higher order multiplets are observed and strongly asymmetric cases are frequent. The oblique magnetic rotator models requires a strong dipole magnetic field which was not detected in RR~Lyrae stars \citep{chadid_mag,kb09}.

Some other ideas were also proposed recently, but lack theoretical elaboration -- for a brief summary see \cite{kovacs_cokon}.
 
 The discovery o period doubling effect in modulated RRab stars triggered theoretical investigations. \cite{kms11} showed that the effect is caused by the half-integer, 9:2 resonance between the fundamental mode and the radial ninth overtone. As analysed by \cite{92resonance}, using the amplitude equations formalism, the same resonance may cause periodic modulation of pulsation. Although period doubling can be reproduced in hydrodynamic models of RR~Lyrae stars, periodic modulation of pulsation can not. It does not invalidate the resonance model behind the Blazhko modulation, as current hydrodynamic models might be too simple to reproduce the effect. Modelling of the high-order, ninth radial overtone, which is a surface mode, with most of its energy trapped in the outer envelope, above the hydrogen ionization front, might be a too challenging task for simple 1D, lagrangian codes that are used in the modelling. In this context we note that \cite{sm12} reproduced both period doubling and periodic modulation of pulsation in hydrodynamic BL Her type models, in which other, lower-order half-integer resonance is in action, the 3:2 resonance between the fundamental mode and radial first overtone. It demonstrates that the resonance mechanism is a plausible model for Blazhko modulation. At the moment, the 9:2 resonance model is the most promising one.

 Whether the mechanism behind the Blazhko modulation is the same in RRab and in RRc stars remains an open questions. Modulation properties in the two groups are clearly different as we have analysed in the preceding sections. The half-integer resonance can in principle work in case of modulated RRc stars, but the problem was not yet investigated in detail, i.e. no particular resonance, which must involve radial first overtone, was proposed so far. Moreover, period doubling was not detected in modulated RRc stars so far; our search for this effect also yielded negative result. Admittedly, pulsation periods of RR Lyrae stars make the detection of period doubling effect in the ground-based observations difficult. Analysis of continuous space photometry for large number of modulated RRc stars might provide a definite answer, whether period doubling exists in modulated RRc stars.

\section{Conclusions}\label{sec.conslusions}
We analysed all first overtone RR~Lyrae stars observed in the Galactic bulge by the OGLE project. We used six seasons of the fourth phase of the project. For some stars we used previous seasons (OGLE-III) if these were available. For some stars it was possible to use data collected over last 20 years. The input sample for the analysis consisted of $10\,826$ stars. The most important results of our study can be summarized as follows.

\begin{itemize}

\item We detected Blazhko effect in 607 stars, which constitutes $5.6\pm0.2$\,per cent of the sample. Among them, 463 were classified as BLm stars, i.e. either triplets or multiplets were detected in their power spectra. 144 stars were classified as BL1 stars, which means that in their power spectra we found only dublet structures. 

\item Both BLm and BL1 structures in the frequency spectra are considered as due to modulation of pulsation. The relative strength of phase and amplitude modulations, and phase relation between the two, determine the amplitudes of the modulation sidepeaks.

\item We selected 105 stars as Blazhko candidates. These stars can be either long-period Blazhko stars or stars in which period change is mimicking quasi-periodic modulation. Due to insufficient data length we cannot classify these stars as firm Blazhko stars. Further monitoring is needed.

\item The shortest Blazhko period is 2.23\,d and it was detected in RRLYR-24030. This is significantly shorter than the shortest modulation periods reported before (around 5\,d). We detected modulation period below 5 days in 13 stars from our sample. Most of them are BL1 stars.

\item The longest Blazhko period we detect is 2954\,d in RRLYR-02478. In some stars we observe period changes on long time scales, that appear regular. These also might be Blazhko stars, but the data length is too short to resolve the modulation sidepeaks.

\item We detected multiperiodic modulation in 47 stars from our sample, which constitutes 8\,per cent of the Blazhko sample. In 43 stars we detected two modulation periods, in three stars we found three modulation periods and in one star we found four modulation periods. The latter case is dubious however as discussed in more detail in Sect.~\ref{ssec:fourmod}. We did not detect any particular pattern in the Petersen diagram, i.e. grouping of the ratios of the modulation periods.

\item We detected multiplet (quintuplet, septuplet) structures in 85 stars, which corresponds to 14\,per cent of the Blazhko stars. Sometimes, the amplitudes of the sidepeaks are strongly asymmetric. In extreme cases we detect modulation sidepeaks only on one side of the radial mode frequency.

\item For stars in which modulation sidepeaks were detected on either side of the radial mode frequency, we investigated the asymmetry in their amplitudes using the asymmetry parameter, $Q$. The distribution is asymmetric with majority of stars having negative $Q$ parameter. The mean value is $-0.1$. In stars with sidepeaks on one side of the radial mode frequency, we equally often observe sidepeaks on higher and on lower frequency side of the first overtone frequency.

\item In 175 Blazhko stars (29\, per cent of the Blazhko sample) we detected a peak in the low-frequency range, at the modulation frequency. It corresponds to modulation of the mean stellar brightness.

\item We detected additional signals in 104 Blazhko stars in our sample, which is 17\,per cent. Four stars fall close to the RRd sequence. 18 stars fall into the $RR_{0.61}$ group. This shows that the Blazhko modulation and non-radial modes might exist in the same object. 24 stars fall into the $RR_{0.68}$ group. Excitation of additional modes and Blazhko modulation are not mutually exclusive phenomena.

\item Using the same sample of stars, \cite{rrabbl} determined the incidence rate of the Blazhko effect for a selected, high-quality sample of fundamental mode pulsators ($40.3$\,per cent). Using similar criteria for RRc stars, we arrive at the incidence rate of the Blazhko effect of $7.6$\,per cent. Thus, in the Galactic bulge, the Blazhko effect is more than 5 times less frequent in RRc stars than in RRab stars.

\end{itemize}

\section*{Acknowledgments}
This research is supported in by the Polish Ministry of Science and Higher Education under grant 0192/DIA/2016/45 within the Diamond Grant Programme for years 2016--2020 (HN). RS and IS are supported by the National Science Center, Poland, grant agreement DEC-2015/17/B/ST9/03421. IS is also supported by the MAESTRO grant 2016/22/A/ST9/00009. The OGLE project has received funding from the National Science Centre, Poland, grant MAESTRO 2014/14/A/ST9/00121 to AU.

\newpage
\appendix
\section[]{List of Blazhko stars}
\begin{table*} 
\centering
\begin{minipage}{190mm}
\caption{Properties of Blazhko stars}
\label{tab.list}
\begin{tabular}{@{}lcccccccl@{}}
\hline
\multicolumn{9}{l}{a - $f_{\rm mod}$ detected in the power spectrum; b - stars with multiplet structures in the power spectrum;}\\ 
\multicolumn{9}{l}{c - stars with additional significant signal; d - multiperiodic Blazhko effect; e - subharmonic of the Blazhko sidepeaks} \\ 
\hline
\hline
Name & type & $P_{\rm 1O}$\thinspace[d] & $P_{\rm BL}$\thinspace[d] &  $A_{\rm 1O}$\thinspace[mag] & $A_+$\thinspace[mag] & $A_-$\thinspace[mag] & Q parameter &  Remarks\\ 
\hline
OGLE-BLG-RRLYR-00065 & BLm & 0.374617(2) & 247(17) & 0.117(3) & 0.025(3) & 0.021(3) & 0.08 &  \\ 
OGLE-BLG-RRLYR-00156 & BLm & 0.379005(2) & 40.15(7) & 0.116(2) & 0.014(2) & 0.021(2) & $-$0.19 &  \\ 
OGLE-BLG-RRLYR-00198 & BLm & 0.3179268(2) & 24.634(2) & 0.0926(7) & 0.0185(7) & 0.0415(7) & $-$0.38 & a \\ 
OGLE-BLG-RRLYR-00416 & BLm & 0.3199298(1) & 2322(65) & 0.0863(6) & 0.0059(7) & 0.0059(7) & 0.00 &  \\ 
OGLE-BLG-RRLYR-00481 & BLm & 0.2652629(1) & 11.500(1) & 0.1127(8) & 0.0246(8) & - & - & a,b \\
\hline
\end{tabular}
\end{minipage}
\end{table*}
\begin{table*} 
\centering
\begin{minipage}{190mm}
\begin{tabular}{@{}lcccccccl@{}}
\hline
Name & type & $P_{\rm 1O}$\thinspace[d] & $P_{\rm BL}$\thinspace[d] &  $A_{\rm 1O}$\thinspace[mag] & $A_+$\thinspace[mag] & $A_-$\thinspace[mag] & Q parameter &  Remarks\\ 
\hline
OGLE-BLG-RRLYR-03808 & BLm & 0.2624569(2) & 26.675(3) & 0.117(1) & 0.055(1) & 0.039(1) & 0.17 &  \\ 
OGLE-BLG-RRLYR-03825 & BLm & 0.27741104(3) & 16.4704(4) & 0.1336(2) & 0.0210(3) & 0.0410(3) & $-$0.32 & a,b,d \\ 
 &  & 0.27741104(3) & 17.304(1) & 0.1336(2) & 0.0105(3) & 0.0140(3) & $-$0.14 &  \\ 
OGLE-BLG-RRLYR-03894 & BL1 & 0.2615433(2) & 7.2938(3) & 0.0779(9) & 0.0488(9) & - & - & a \\ 
OGLE-BLG-RRLYR-03963 & BLm & 0.30115522(10) & 71.25(1) & 0.1182(7) & 0.0536(6) & 0.0526(7) & 0.01 & b \\ 
OGLE-BLG-RRLYR-04033 & BL1 & 0.2690522(2) & 8.2824(3) & 0.0925(9) & 0.0496(8) & - & - & a \\ 
OGLE-BLG-RRLYR-04039 & BLm & 0.26697593(4) & 1691(20) & 0.0898(3) & 0.0033(3) & 0.0052(3) & $-$0.22 &  \\
\hline
\end{tabular}
\end{minipage}
\end{table*}
\begin{table*} 
\centering
\begin{minipage}{190mm}
\begin{tabular}{@{}lcccccccl@{}}
\hline
Name & type & $P_{\rm 1O}$\thinspace[d] & $P_{\rm BL}$\thinspace[d] &  $A_{\rm 1O}$\thinspace[mag] & $A_+$\thinspace[mag] & $A_-$\thinspace[mag] & Q parameter &  Remarks\\ 
\hline
OGLE-BLG-RRLYR-07072 & BLm & 0.2986516(2) & 688(4) & 0.0868(8) & 0.0182(9) & 0.0210(9) & $-$0.07 &  \\ 
OGLE-BLG-RRLYR-07085 & BLm & 0.29191494(7) & 1915(15) & 0.1188(10) & 0.018(1) & 0.020(1) & $-$0.05 & c \\ 
OGLE-BLG-RRLYR-07114 & BL1 & 0.3050536(3) & 7.4274(8) & 0.112(1) & - & 0.021(1) & - &  \\ 
OGLE-BLG-RRLYR-07117 & BLm & 0.2747862(1) & 161.09(9) & 0.0899(6) & 0.0435(6) & - & - & a,b \\ 
OGLE-BLG-RRLYR-07133 & BLm & 0.25760270(7) & 21.511(2) & 0.0751(3) & 0.0139(3) & 0.0204(3) & $-$0.19 &  \\ 
OGLE-BLG-RRLYR-07173 & BLm & 0.26349513(6) & 195.93(5) & 0.0853(6) & 0.0164(6) & 0.0641(6) & $-$0.59 & a \\ 
OGLE-BLG-RRLYR-07327 & BLm & 0.2627767(1) & 16.0256(8) & 0.0636(5) & - & 0.0579(5) & - & a,b \\
\hline
\end{tabular}
\end{minipage}
\end{table*}
\begin{table*} 
\centering
\begin{minipage}{190mm}
\begin{tabular}{@{}lcccccccl@{}}
\hline
Name & type & $P_{\rm 1O}$\thinspace[d] & $P_{\rm BL}$\thinspace[d] &  $A_{\rm 1O}$\thinspace[mag] & $A_+$\thinspace[mag] & $A_-$\thinspace[mag] & Q parameter &  Remarks\\ 
\hline
OGLE-BLG-RRLYR-09482 & BLm & 0.25387053(6) & 4.91735(4) & 0.0635(2) & 0.0464(2) & - & - & a \\ 
OGLE-BLG-RRLYR-09542 & BLm & 0.2614356(1) & 24.671(2) & 0.0733(4) & - & 0.0449(4) & - & a,b \\ 
OGLE-BLG-RRLYR-09556 & BLm & 0.26774872(8) & 257.3(2) & 0.1002(4) & 0.0147(4) & 0.0594(4) & $-$0.60 & a \\ 
OGLE-BLG-RRLYR-09605 & BLm & 0.25572062(8) & 24.607(4) & 0.1004(4) & 0.0156(4) & 0.0135(4) & 0.07 & c \\ 
OGLE-BLG-RRLYR-09609 & BLm & 0.27688862(4) & 7.1590(7) & 0.1268(2) & 0.0021(3) & 0.0041(2) & $-$0.33 &  \\ 
OGLE-BLG-RRLYR-09697 & BL1 & 0.26682288(9) & 41.86(1) & 0.1345(7) & - & 0.0302(7) & - &  \\ 
OGLE-BLG-RRLYR-09706 & BLm & 0.25252704(4) & 47.915(3) & 0.0561(1) & 0.0284(1) & - & - & a,b \\
\hline
\end{tabular}
\end{minipage}
\end{table*}
\begin{table*} 
\centering
\begin{minipage}{190mm}
\begin{tabular}{@{}lcccccccl@{}}
\hline
Name & type & $P_{\rm 1O}$\thinspace[d] & $P_{\rm BL}$\thinspace[d] &  $A_{\rm 1O}$\thinspace[mag] & $A_+$\thinspace[mag] & $A_-$\thinspace[mag] & Q parameter &  Remarks\\ 
\hline
OGLE-BLG-RRLYR-12040 & BL1 & 0.25134710(4) & 8.5430(1) & 0.0749(1) & 0.0237(1) & - & - & a \\ 
OGLE-BLG-RRLYR-12072 & BL1 & 0.2569269(1) & 41.103(5) & 0.0674(4) & - & 0.0476(4) & - & a \\ 
OGLE-BLG-RRLYR-12135 & BLm & 0.39194372(5) & 33.453(1) & 0.1135(1) & 0.0281(1) & 0.0199(1) & 0.17 & a \\ 
OGLE-BLG-RRLYR-12140 & BLm & 0.25833097(6) & 17.412(1) & 0.0974(3) & 0.0075(3) & 0.0183(3) & $-$0.42 &  \\ 
OGLE-BLG-RRLYR-12164 & BL1 & 0.24684762(2) & 4.8230(1) & 0.0687(2) & 0.0134(2) & - & - & a \\ 
OGLE-BLG-RRLYR-12169 & BLm & 0.25411046(5) & 33.456(3) & 0.1026(3) & 0.0152(3) & 0.0189(3) & $-$0.11 &  \\ 
OGLE-BLG-RRLYR-12182 & BLm & 0.29601537(5) & 2271(31) & 0.1262(4) & 0.0093(4) & 0.0094(4) & $-$0.01 & c \\
\hline
\end{tabular}
\end{minipage}
\end{table*}
\begin{table*} 
\centering
\begin{minipage}{190mm}
\begin{tabular}{@{}lcccccccl@{}}
\hline
Name & type & $P_{\rm 1O}$\thinspace[d] & $P_{\rm BL}$\thinspace[d] &  $A_{\rm 1O}$\thinspace[mag] & $A_+$\thinspace[mag] & $A_-$\thinspace[mag] & Q parameter &  Remarks\\ 
\hline
OGLE-BLG-RRLYR-14358 & BLm & 0.27174232(8) & 12.8432(9) & 0.1115(5) & 0.0192(5) & - & - & b \\ 
OGLE-BLG-RRLYR-14446 & BL1 & 0.25141410(9) & 29.11(2) & 0.1185(6) & - & 0.0090(6) & - &  \\ 
OGLE-BLG-RRLYR-14455 & BLm & 0.25645034(6) & 108.15(2) & 0.0919(4) & 0.0265(4) & 0.0402(4) & $-$0.21 &  \\ 
OGLE-BLG-RRLYR-14487 & BL1 & 0.2579873(2) & 185.3(1) & 0.0741(8) & 0.0592(8) & - & - &  \\ 
OGLE-BLG-RRLYR-14503 & BLm & 0.32638291(9) & 41.124(3) & 0.1304(5) & 0.0871(5) & - & - & a,b \\ 
OGLE-BLG-RRLYR-14542 & BLm & 0.31023376(7) & 28.745(3) & 0.1319(4) & 0.0129(4) & 0.0212(4) & $-$0.25 &  \\ 
OGLE-BLG-RRLYR-14547 & BLm & 0.29877258(5) & 1922(43) & 0.1096(6) & 0.0038(6) & 0.0049(6) & $-$0.13 &  \\
\hline
\end{tabular}
\end{minipage}
\end{table*}
\begin{table*} 
\centering
\begin{minipage}{190mm}
\begin{tabular}{@{}lcccccccl@{}}
\hline
Name & type & $P_{\rm 1O}$\thinspace[d] & $P_{\rm BL}$\thinspace[d] &  $A_{\rm 1O}$\thinspace[mag] & $A_+$\thinspace[mag] & $A_-$\thinspace[mag] & Q parameter &  Remarks\\ 
\hline
OGLE-BLG-RRLYR-17544 & BL1 & 0.3701184(8) & 65.35(6) & 0.077(2) & 0.038(2) & - & - &  \\ 
OGLE-BLG-RRLYR-17688 & BL1 & 0.3859845(3) & 9.603(2) & 0.135(1) & 0.013(1) & - & - &  \\ 
OGLE-BLG-RRLYR-17971 & BL1 & 0.2680346(2) & 12.7815(10) & 0.117(1) & - & 0.044(1) & - & a \\
OGLE-BLG-RRLYR-18001 & BLm & 0.2789929(2) & 12.673(1) & 0.101(1) & 0.035(1) & 0.034(1) & 0.02 &  \\ 
OGLE-BLG-RRLYR-18013 & BL1 & 0.2765035(2) & 12.2010(6) & 0.112(1) & - & 0.038(1) & - &  \\ 
OGLE-BLG-RRLYR-18315 & BL1 & 0.2515577(2) & 255.3(6) & 0.086(1) & - & 0.044(1) & - &  \\ 
OGLE-BLG-RRLYR-18390 & BLm & 0.3423858(3) & 74.66(2) & 0.114(1) & 0.038(1) & 0.082(1) & $-$0.36 & b \\
\hline
\end{tabular}
\end{minipage}
\end{table*}
\begin{table*} 
\centering
\begin{minipage}{190mm}
\begin{tabular}{@{}lcccccccl@{}}
\hline
Name & type & $P_{\rm 1O}$\thinspace[d] & $P_{\rm BL}$\thinspace[d] &  $A_{\rm 1O}$\thinspace[mag] & $A_+$\thinspace[mag] & $A_-$\thinspace[mag] & Q parameter &  Remarks\\ 
\hline
OGLE-BLG-RRLYR-24030 & BLm & 0.2910904(2) & 2.23360(3) & 0.260(3) & 0.117(3) & - & - & a,b \\ 
OGLE-BLG-RRLYR-24085 & BLm & 0.2796914(5) & 69.1(1) & 0.110(3) & 0.027(3) & 0.025(3) & 0.02 &  \\ 
OGLE-BLG-RRLYR-24316 & BLm & 0.3565236(3) & 15.098(2) & 0.0646(7) & 0.0148(7) & 0.0142(7) & 0.02 &  \\ 
OGLE-BLG-RRLYR-24625 & BLm & 0.2498633(1) & 12.902(2) & 0.0983(6) & 0.0108(7) & 0.0106(7) & 0.01 &  \\ 
OGLE-BLG-RRLYR-24702 & BLm & 0.26745744(8) & 22.248(4) & 0.1436(7) & 0.0142(7) & 0.0168(7) & $-$0.08 &  \\ 
OGLE-BLG-RRLYR-24729 & BLm & 0.3310926(6) & 63.68(6) & 0.101(2) & 0.030(2) & 0.044(2) & $-$0.18 & c \\ 
OGLE-BLG-RRLYR-25165 & BLm & 0.3157133(6) & 51.77(5) & 0.084(2) & 0.013(2) & 0.024(2) & $-$0.30 & c \\
\hline
\end{tabular}
\end{minipage}
\end{table*}
\begin{table*} 
\centering
\begin{minipage}{190mm}
\begin{tabular}{@{}lcccccccl@{}}
\hline
Name & type & $P_{\rm 1O}$\thinspace[d] & $P_{\rm BL}$\thinspace[d] &  $A_{\rm 1O}$\thinspace[mag] & $A_+$\thinspace[mag] & $A_-$\thinspace[mag] & Q parameter &  Remarks\\ 
\hline
OGLE-BLG-RRLYR-31374 & BL1 & 0.31427358(5) & 25.871(2) & 0.1103(2) & 0.0159(2) & - & - & a,c \\ 
OGLE-BLG-RRLYR-31383 & BLm & 0.3199999(1) & 16.132(1) & 0.1236(6) & 0.0264(6) & 0.0222(6) & 0.09 &  \\ 
OGLE-BLG-RRLYR-31557 & BLm & 0.26052206(7) & 58.90(5) & 0.0637(2) & 0.0037(2) & - & - & e \\ 
OGLE-BLG-RRLYR-31569 & BL1 & 0.24595110(10) & 39.39(5) & 0.0712(4) & 0.0022(4) & - & - &  \\ 
OGLE-BLG-RRLYR-31577 & BLm & 0.3532671(4) & 17.463(2) & 0.0596(7) & 0.0525(7) & - & - & a,b \\ 
OGLE-BLG-RRLYR-31615 & BLm & 0.2734660(2) & 8.886(1) & 0.141(2) & 0.017(2) & 0.017(2) & 0.00&  \\ 
OGLE-BLG-RRLYR-31643 & BLm & 0.2826973(2) & 53.36(2) & 0.1129(9) & 0.0156(9) & 0.0281(9) & $-$0.29 &  \\
\hline
\end{tabular}
\end{minipage}
\end{table*}
\begin{table*} 
\centering
\begin{minipage}{190mm}
\begin{tabular}{@{}lcccccccl@{}}
\hline
Name & type & $P_{\rm 1O}$\thinspace[d] & $P_{\rm BL}$\thinspace[d] &  $A_{\rm 1O}$\thinspace[mag] & $A_+$\thinspace[mag] & $A_-$\thinspace[mag] & Q parameter &  Remarks\\ 
\hline
OGLE-BLG-RRLYR-33819 & BL1 & 0.3047738(2) & 11.0283(3) & 0.0742(4) & - & 0.0557(4) & - & a \\ 
OGLE-BLG-RRLYR-33862 & BLm & 0.2194080(1) & 20.585(2) & 0.0609(6) & 0.0147(5) & 0.0226(5) & $-$0.21 & a,c,d \\ 
 &  & 0.2194080(1) & 17.510(3) & 0.0609(6) & 0.0026(5) & 0.0136(5) & -0.68 &  \\ 
OGLE-BLG-RRLYR-33918 & BL1 & 0.2585291(1) & 377.6(8) & 0.0931(6) & 0.0369(7) & - & - & a,d \\ 
 &  & 0.2585291(1) & 41.75(1) & 0.0931(6) & 0.0178(5) & - & - &  \\ 
OGLE-BLG-RRLYR-33932 & BLm & 0.3393862(2) & 10.9347(8) & 0.0866(5) & 0.0131(5) & 0.0165(5) & $-$0.11 &  \\ 
OGLE-BLG-RRLYR-33964 & BL1 & 0.21544331(9) & 7.3079(6) & 0.0502(3) & 0.0086(3) & - & - & c \\
\hline
\end{tabular}
\end{minipage}
\end{table*}
\begin{table*} 
\centering
\begin{minipage}{190mm}
\begin{tabular}{@{}lcccccccl@{}}
\hline
Name & type & $P_{\rm 1O}$\thinspace[d] & $P_{\rm BL}$\thinspace[d] &  $A_{\rm 1O}$\thinspace[mag] & $A_+$\thinspace[mag] & $A_-$\thinspace[mag] & Q parameter &  Remarks\\ 
\hline
OGLE-BLG-RRLYR-38198 & BLm & 0.3590991(5) & 12.395(2) & 0.0928(8) & 0.0220(8) & 0.0324(7) & $-$0.19 &  \\ 
OGLE-BLG-RRLYR-38244 & BL1 & 0.2728477(7) & 11.570(2) & 0.075(1) & - & 0.059(1) & - & a \\ 
OGLE-BLG-RRLYR-38258 & BLm & 0.277797(1) & 9.904(4) & 0.083(3) & 0.019(3) & 0.022(3) & $-$0.08 &  \\ 
OGLE-BLG-RRLYR-38288 & BLm & 0.280621(1) & 50.84(9) & 0.097(3) & 0.031(3) & 0.034(3) & $-$0.06 &  \\ 
\hline
\end{tabular}
\end{minipage}
\end{table*}
\section[]{List of candidates for Blazhko stars}
\begin{table*} 
\centering
\begin{minipage}{190mm}
\caption{Properties of Blazhko candidates}
\label{tab.list}
\begin{tabular}{@{}lccccccc@{}}
\hline
\hline
Name & type & $P_{\rm 1O}$\thinspace[d] & $P_{\rm BL}$\thinspace[d] &  $A_{\rm 1O}$\thinspace[mag] & $A_+$\thinspace[mag] & $A_-$\thinspace[mag] & Q parameter \\ 
\hline
OGLE-BLG-RRLYR-00200 & BLm & 0.3197446(3) & 1235(84) & 0.128(1) & 0.014(2) & 0.014(2) & 0.00 \\
OGLE-BLG-RRLYR-00250 & BLm & 0.26927011(10) & 2235(30) & 0.0887(8) & 0.0131(9) & 0.0128(10) & 0.01 \\
OGLE-BLG-RRLYR-00436 & BLm & 0.3881490(1) & 1269(8) & 0.138(1) & 0.018(1) & 0.016(1) & 0.05 \\
OGLE-BLG-RRLYR-00925 & BLm & 0.37111253(7) & 1207(10) & 0.165(1) & 0.010(1) & 0.007(1) & 0.17 \\
OGLE-BLG-RRLYR-00948 & BLm & 0.3897224(1) & 1920(35) & 0.1175(6) & 0.0070(7) & 0.0082(7) & $-$0.08 \\
OGLE-BLG-RRLYR-01290 & BLm & 0.3539806(3) & 2177(11) & 0.1345(9) & 0.0213(9) & 0.0206(9) & 0.02 \\
 &  & 0.3539806(3) & 698(6) & 0.1345(9) & 0.008(1) & 0.0076(9) & 0.02 \\
\hline
\end{tabular}
\end{minipage}
\end{table*}
\begin{table*} 
\centering
\begin{minipage}{190mm}
\begin{tabular}{@{}lccccccc@{}}
\hline
Name & type & $P_{\rm 1O}$\thinspace[d] & $P_{\rm BL}$\thinspace[d] &  $A_{\rm 1O}$\thinspace[mag] & $A_+$\thinspace[mag] & $A_-$\thinspace[mag] & Q parameter \\ 
\hline
OGLE-BLG-RRLYR-10761 & BLm & 0.31271821(6) & 2114(47) & 0.1229(6) & 0.0045(6) & 0.0060(6) & $-$0.15 \\
OGLE-BLG-RRLYR-11184 & BLm & 0.27981642(5) & 1868(21) & 0.1516(9) & 0.0120(9) & 0.0111(9) & 0.04 \\
OGLE-BLG-RRLYR-11186 & BLm & 0.30101432(4) & 2683(45) & 0.1315(7) & 0.0066(6) & 0.0060(6) & 0.05 \\
OGLE-BLG-RRLYR-11290 & BLm & 0.29310793(3) & 1000(8) & 0.1481(6) & 0.0060(6) & 0.0051(6) & 0.08 \\
OGLE-BLG-RRLYR-11308 & BLm & 0.25384758(3) & 1280(19) & 0.1179(5) & 0.0034(5) & 0.0044(5) & $-$0.13 \\
OGLE-BLG-RRLYR-11432 & BLm & 0.3477461(2) & 1719(27) & 0.1249(5) & 0.0058(5) & 0.0049(5) & 0.08 \\
OGLE-BLG-RRLYR-11504 & BLm & 0.30098354(7) & 2022(72) & 0.1239(6) & 0.0038(6) & 0.0041(6) & $-$0.04 \\
\hline
\end{tabular}
\end{minipage}
\end{table*}
\section[]{Stars with additional frequencies.}

 \begin{table*}
  \begin{minipage}{150mm}
   \caption{Stars with additional frequencies in power spectrum.}
   \label{tab.a2}
   \begin{tabular}{@{}||lcccccl@{}}
 
 ID& $P_{\rm 1O}$& $P_{\rm X}$&$P_S/P_L$ &${A_{\rm 1O}}$&$A_{\rm X}/A_{\rm 1O}$& Possible explanation\\
  \hline
OGLE-BLG-RRLYR-00719 & 0.26569 & 0.26319 & 0.99059 & 0.11564 & 0.0605 & nr/BL \\
OGLE-BLG-RRLYR-00959 & 0.27026 & 0.26688 & 0.98747 & 0.08773 & 0.1281 & nr/BL \\
 & 0.27026 & 0.56903 & 0.47496 & 0.08773 & 0.0527 &  \\
 & 0.27026 & 0.13918 & 0.51497 & 0.08773 & 0.0567 &  \\
OGLE-BLG-RRLYR-01283 & 0.31583 & 0.31109 & 0.98497 & 0.10217 & 0.0478 & nr/BL \\
OGLE-BLG-RRLYR-03053 & 0.28984 & 0.42349 & 0.68441 & 0.12790 & 0.0500 & $RR_{0.68}$ \\
OGLE-BLG-RRLYR-03058 & 0.38629 & 0.56392 & 0.68501 & 0.13846 & 0.0199 & $RR_{0.68}$ \\
\end{tabular}
 \end{minipage}
 \end{table*}

   \begin{table*} 
  \centering
   \begin{minipage}{150mm}
 
   \begin{tabular}{@{}||lcccccl@{}}
 
 ID& $P_{\rm 1O}$& $P_{\rm X}$&$P_S/P_L$ &${A_{\rm 1O}}$&$A_{\rm 1O}/A_{\rm X}$& Possible explanation\\
  \hline
  OGLE-BLG-RRLYR-13148 & 0.31095 & 0.45412 & 0.68472 & 0.12139 & 0.0309 & $RR_{0.68}$ \\
 & 0.31095 & 0.30390 & 0.97736 & 0.12139 & 0.0135 & nr/BL \\
OGLE-BLG-RRLYR-13161 & 0.25376 & 0.13471 & 0.53085 & 0.05128 & 0.0591 &  \\
OGLE-BLG-RRLYR-13253 & 0.32590 & 0.20555 & 0.63073 & 0.12000 & 0.0211 & $RR_{0.61}$ \\
OGLE-BLG-RRLYR-13280 & 0.30259 & 0.32421 & 0.93332 & 0.04035 & 0.1435 & nr/BL \\
OGLE-BLG-RRLYR-13295 & 0.29307 & 0.30204 & 0.97030 & 0.08517 & 0.0673 & nr/BL \\
OGLE-BLG-RRLYR-13328 & 0.31801 & 0.19491 & 0.61291 & 0.11774 & 0.0199 & $RR_{0.61}$ \\
\end{tabular}
 \end{minipage}
 \end{table*}
 
    \begin{table*} 
  \centering
   \begin{minipage}{150mm}
 
   \begin{tabular}{@{}||lcccccl@{}}
 
 ID& $P_{\rm 1O}$& $P_{\rm X}$&$P_S/P_L$ &${A_{\rm 1O}}$&$A_{\rm 1O}/A_{\rm X}$& Possible explanation\\
  \hline
  OGLE-BLG-RRLYR-34540 & 0.27718 & 0.40699 & 0.68106 & 0.13909 & 0.0376 & $RR_{0.68}$ \\
OGLE-BLG-RRLYR-35091 & 0.25910 & 0.35881 & 0.72210 & 0.08037 & 0.0793 &  \\
OGLE-BLG-RRLYR-35328 & 0.30681 & 0.44596 & 0.68799 & 0.12569 & 0.0244 & $RR_{0.68}$ \\
OGLE-BLG-RRLYR-35820 & 0.23864 & 0.14641 & 0.61355 & 0.11420 & 0.0459 & $RR_{0.61}$ \\
OGLE-BLG-RRLYR-36049 & 0.35747 & 0.52181 & 0.68506 & 0.08937 & 0.0705 & $RR_{0.68}$ \\
OGLE-BLG-RRLYR-36334 & 0.28046 & 0.27782 & 0.99060 & 0.11352 & 0.0926 & nr/BL \\
 \end{tabular}
 \end{minipage}
 \end{table*}

\bsp

\label{lastpage}

\end{document}